\documentclass[final,leqno]{siamltex1213}
\usepackage{graphicx}
\usepackage{epstopdf}
\usepackage{fancyhdr}
\usepackage{paralist}
\usepackage{subcaption}
\usepackage{amsmath}
\usepackage{amssymb}
\usepackage{algorithm}
\usepackage{algcompatible}
\usepackage{xcolor}
\usepackage{listings}
\usepackage{xspace}
\usepackage{soul} 
\setstcolor{red}

\newcommand{\code}[1]{\texttt{#1}\xspace}
\newcommand{\tbx}{\ensuremath{\mathbf{\tilde x}}\xspace}
\newcommand{\tbr}{\ensuremath{\mathbf{\tilde r}}\xspace}
\newcommand{\epsm}{\ensuremath{\epsilon_{\text{mach}}}\xspace}
\newcommand{\bx}{\ensuremath{\mathbf x}\xspace}

\newcommand{\tlambda}{\ensuremath{\tilde\lambda}\xspace}

\newcommand{\tbu}{\ensuremath{\mathbf {\tilde u}}\xspace}
\newcommand{\tbv}{\ensuremath{\mathbf {\tilde v}}\xspace}

\newcommand{\bbv}{\ensuremath{\mathbf {\bar v}}\xspace}
\newcommand{\bu}{\ensuremath{\mathbf u}\xspace}
\newcommand{\bv}{\ensuremath{\mathbf v}\xspace}

\newcommand{\by}{\ensuremath{\mathbf y}\xspace}

\newcommand{\bo}{\ensuremath{\mathbf 0}\xspace}

\newcommand{\tsigma}{\ensuremath{\tilde\sigma}\xspace}

\title{PRIMME\_SVDS: A High-Performance Preconditioned SVD Solver for Accurate Large-Scale Computations}
\author{Lingfei Wu\footnotemark[1],
   Eloy Romero\footnotemark[1],
   Andreas Stathopoulos\footnotemark[1]}

\begin{document}
\maketitle
\renewcommand{\thefootnote}{\fnsymbol{footnote}}
\footnotetext[1]{Department of Computer Science, College of William and Mary, Williamsburg, Virginia 23187-8795, 
U.S.A. (lwu@email.wm.edu, eloy@cs.wm.edu, andreas@cs.wm.edu)} 
\renewcommand{\thefootnote}{\arabic{footnote}}

\begin{abstract}
The increasing number of applications requiring the solution of large scale singular value problems has rekindled an interest in iterative methods for the SVD. 
Some promising recent advances in large scale iterative methods are still plagued by slow convergence and accuracy limitations for computing smallest singular triplets. Furthermore, their current implementations in MATLAB cannot address the required large problems. Recently, we presented a preconditioned, two-stage method to effectively and accurately compute a small number of extreme singular triplets. In this research, we present a high-performance library, PRIMME\_SVDS, that implements our hybrid method based on the state-of-the-art eigensolver package PRIMME for both largest and smallest singular values. PRIMME\_SVDS fills a gap in production level software for computing the partial SVD, especially with preconditioning. The numerical experiments demonstrate its superior performance compared to other state-of-the-art software and its good parallel performance under strong and weak scaling.
\end{abstract}

\section{Introduction}

We consider the problem of finding a small number, $k\ll n$, of extreme singular values and corresponding left and right singular vectors of a large sparse matrix $ A \in \mathbb{R}^{m \times n} $ ($ m \geq n $). These singular values and vectors satisfy,
\begin{equation}
Av_i = \sigma_i u_i, 
\end{equation}
where the singular values are labeled in ascending order, $\sigma_1 \leq \sigma_2 \leq \ldots \leq \sigma_n$. The matrices $ U = [u_{1},\ldots , u_{n}] \in \mathbb{R}^{m \times n}$ and $ V = [v_{1},\ldots , v_{n}] \in \mathbb{R}^{n \times n}$ have orthonormal columns. $ (\sigma_{i}, u_{i}, v_{i} )$ is called a singular triplet of $A$. The singular triplets corresponding to largest (smallest) singular values are referred as largest (smallest) singular triplets. The Singular Value Decomposition (SVD) can always be computed \cite{golub1965calculating}. 

The SVD is one of the most widely used computational kernels in various scientific and engineering areas. The computation of a few of the largest singular triplets plays a critical role in machine learning for computing a low-rank matrix approximation \cite{williams2001using, wuyenrevisiting, chenwuefficient} and the nuclear norm \cite{recht2010guaranteed}, in data mining for latent semantic indexing \cite{deerwester1990indexing}, in statistics for principal component analysis \cite{jolliffe2002principal}, and in signal processing and pattern recognition as an important filtering tool. 
Calculating a few smallest singular triplets plays an important role in linear algebra applications such as computing pseudospectra, and determining the range, null space, and rank of a matrix \cite{Golub1996MC, trefethen1999computation}, in machine learning for the total least squares problems \cite{golub1980analysis}, and as a general deflation mechanism, e.g., for computing the trace of the matrix inverse \cite{wu2015estimating}. 
 
For large scale problems, memory demands and computational complexity necessitate
the use of iterative methods. 
Over the last two decades, iterative methods for the SVD have been
developed \cite{golub1965calculating, philippe1997computation,
kokiopoulou2004computing, cullum1983Lanczos, dongarra1983improving,
baglama2006restarted, baglama2005augmented, jia2003implicitly, jia2010refined,
hochstenbach2001jacobi, Wu2015PHSVD} to effectively compute a few singular
triplets under limited memory.  In particular, the computation of the smallest
singular triplets presents challenges both to the speed of convergence and to the
accuracy of iterative methods, even for problems with moderate conditioning.
Many recent research efforts attempt to address this challenge with new or modified
  iterative methods
\cite{baglama2006restarted, baglama2005augmented, jia2003implicitly,
jia2010refined, hochstenbach2001jacobi}. 
These methods still lack the necessary robustness and display irregular or slow 
  convergence in cases of clustered spectra.
Most importantly, they are only available in MATLAB research implementations that 
  cannot solve large-scale problems.
In our previous work \cite{Wu2015PHSVD}, we presented a hybrid, two-stage method that achieves both efficiency and accuracy for both largest and smallest singular values under limited memory. In addition, the method can take full advantage of preconditioning to significantly accelerate the computation, which is the key towards large-scale SVD computation for certain real-world problems. Our previous results showed substantial improvements in efficiency and robustness over other methods.

With all this algorithmic activity it is surprising that there is a lack of good quality software for computing the partial SVD, especially with preconditioning. SVDPACK \cite{berry1992large} and PROPACK \cite{larsen1998Lanczos} can efficiently compute largest singular triplets but they are either single-threaded or only support shared-memory parallel computing. In addition, they are slow and unreliable for computing the smallest singular triplets and cannot directly take advantage of preconditioning. 
SLEPc offers a more robust implementation of the thick restarted Lanczos 
bidiagonalization method (LBD) \cite{andez2008robust} which can be used in both 
distributed and shared memory environments. SLEPc also offers some limited 
functionality for computing the partial SVD through its eigensolvers \cite{hernandez2005slepc}. 
However, neither approach is tuned for accurate computation of smallest singular triplets,
  and only the eigensolver approach can use preconditioning which limits the desired
  accuracy or the practical efficiency of the package.
There is a clear need for high quality, high performance SVD software that addresses 
  the above shortcomings of current software and that provides a flexible interface suitable 
  for both black-box and advanced usage.

In this work we address this need by developing a high quality SVD library, PRIMME\_SVDS, based on the state-of-the-art package PRIMME (PReconditioned Iterative MultiMethod Eigensolver \cite{stathopoulos2010primme}). We highlight three main contributions:
\begin{itemize}
    \item We present the modifications and functionality extensions of 
	    PRIMME that are required to implement our two stage PHSVDS 
	    method \cite{Wu2015PHSVD} so that it achieves full accuracy and 
	    high-performance, for both smallest and largest singular values 
	    and with or without preconditioning. 
    \item We provide intuitive user interfaces in C, MATLAB, Python, and R 
	    for both ordinary and advanced users to fully exploit the power 
	    of PRIMME\_SVDS. We also discuss distributed and shared memory 
	    interfaces.
    \item We demonstrate with numerical experiments on large scale matrices
	    that PRIMME\_SVDS is more efficient and significantly more robust 
	    than the two most widely used software packages, PROPACK and SLEPc, 
		even without a preconditioner. 
	    We also demonstrate its good parallel scalability.
\end{itemize}
Since this is a software paper, the presentation of our method is given 
  in a more abstract way that relates to the implementation. 
  Further theoretical and algorithmic details can be found in \cite{Wu2015PHSVD}.

\section{Related work}\label{sec:related work}
The partial SVD problem can be solved as a Hermitian eigenvalue problem 
  or directly using bidiagonalization methods \cite{Golub1996MC}.
We introduce these approaches and discuss the resulting 
state-of-the-art SVD methods. Next, we review the state-of-practice software,
both eigensolver and dedicated SVD software.

\subsection{Approaches and Methods for Computing SVD Problems}
\label{subsec:svd method}
The first approach computes eigenpairs of the normal equations matrix 
$C = A^T A \in \mathbb C^{n \times n}$ 
(or $AA^T \in \mathbb C^{m \times m}$ if $m<n$).
The eigenpairs of $C$ correspond to the squares of the singular values 
  and the right singular vectors of $A$,
 $(A^TA)V = V\texttt{diag}\left(\sigma_1^2, \ldots, \sigma_n^2\right)$.
If $\sigma_i \neq 0$, the corresponding left singular vectors are obtained as $u_i = \frac{1}{\sigma_i} A v_i$. 
Although extreme Hermitian eigenvalue problems can be solved very efficiently,
  $C$ can be ill-conditioned.
Assume $(\tsigma_i,\tbu_i,\tbv_i)$ is a computed singular triplet
  corresponding to the exact triplet $(\sigma_i,\bu_i,\bv_i)$ 
  with residual satisfying $\|C\tbv_i - \tbv_i \tsigma_i\|\leq \|A\|^2 \epsm$.
Then the relative errors satisfy the relations \cite{stewart2001matrix},
\begin{eqnarray}
|\frac{\tsigma_i - \sigma_i}{\sigma_i}| & \leq & \frac{\sigma_n^2}{\sigma_i^2} \epsm \equiv \kappa_i^2 \epsm, \label{eq:value-error} \\
\sin \angle(\tbv_i,\bv_i) & \lesssim & \frac{\sqrt 2 \epsm} {|\kappa_i^{-1}-\kappa_j^{-1}| |\kappa_i^{-1}+\kappa_j^{-1}|}, \label{eq:v-error} \\
\frac{\|\tbu_i - \bu_i\|_2}{\|\bu_i\|_2} & \lesssim & \kappa_i ( \|\tbv_i - \bv_i\|_2 + \gamma \epsm), \label{eq:u-error} 
\end{eqnarray} 
where $\sigma_j$ is the closest singular value to $\sigma_i$, 
$\gamma$ is a constant that depends on the dimensions of $A$, 
  and $\epsm$ is the machine precision. 
Since all singular values (except the largest $\sigma_n$) will
  lose accuracy, this approach is typically followed by a second stage of iterative refinement.
The singular triplets can be refined one by one \cite{philippe1997computation,dongarra1983improving,berry1992large} or collectively as an initial subspace for an iterative eigensolver \cite{Wu2015PHSVD}.
This is especially needed when seeking smallest singular values.

The second approach seeks eigenpairs of the augmented matrix 
  $ B = \left[\begin{array}{cc} 0 & A^T \\ A & 0\end{array}\right]$, 
where $B \in \mathbb C^{(m+n) \times (m+n)}$.
If $U^{\perp}$ is a basis for the orthogonal complement subspace of $U$,
where $U^{\perp} \in \mathbb C^{m \times (m-n)}$, and define the orthonormal matrix
\begin{equation}
    Y = \frac{1}{\sqrt{2}} 
    \left[ \begin{array}{ccc}
         V & -V & 0 \\
         U &  U & \sqrt{2}U^{\perp}
    \end{array} \right],
    \label{eq:B_Initguess}
\end{equation}
then the eigenvalue decomposition of $B$ corresponds to
\cite{golub1965calculating,Golub1996MC}
\begin{equation}
    BY = Y\texttt{diag}\left(\sigma_1, \ldots, \sigma_n, -\sigma_1, \ldots, -\sigma_n, \underbrace{0, \ldots, 0}_{m-n}\right).
    \label{eq:B_Augmented}
\end{equation}
This approach can compute all singular values accurately, i.e., 
with residual norm close to $\|A\| \epsm$ or relative accuracy
$O(\kappa_i\epsm)$.
The main disadvantage of this approach is that when seeking the smallest 
  singular values, the eigenvalue problem is a maximally interior one and thus
  Krylov iterative methods converge slowly and unreliably.
Let $K_l(C,\bv_1)$ be the Krylov subspace of dimension $l$ of matrix $C$ 
starting with initial vector $\bv_1$. It is easy to see that
$K_{2l}(B,\bbv_1) = 
\left[\begin{array}{c} K_{l}(C,\bv_1) \\ \bo \end{array}\right]
\oplus 
\left[\begin{array}{c} \bo \\ AK_{l}(C,\bv_1) \end{array}\right]$, 
where $\bbv_1$ is the vector $[\bv_1; \ \bo]$\ \footnote{We use MATLAB notation for column concatenation, $[\bu;\ \bv]=[\bu^T\ \bv^T]^T$.}. Therefore, an unrestarted Krylov method on $B$ with the proper initial vector and extraction method converges twice slower than that on $C$ \cite{Wu2015PHSVD}.
In practice, restarted methods on $C$ can converge much faster than on $B$, partly because the Rayleigh-Ritz projection for interior eigenvalues is less efficient due to presence of spurious Ritz values \cite{parlett1980symmetric}. Harmonic projection \cite{morgan1991computing,Morgan1998Harmonic} and refined projection \cite{jia1997refined} have been proposed to overcome this difficulty. 
However, near-optimal restarted methods on $C$ still converge significantly faster \cite{Wu2015PHSVD}.

The Lanczos bidiagonalization (LBD) method \cite{Golub1996MC, golub1965calculating} addresses the SVD problem directly, with many variants proposed in the last ten years \cite{larsen2001combining, jia2003implicitly, kokiopoulou2004computing, baglama2005augmented, baglama2006restarted, baglama2013implicitly, jia2010refined}. It is accepted as an accurate and more efficient method for seeking extreme singular triplets since it works on $A$ directly which avoids the numerical problems of squaring. Starting with  initial unit vectors $p_1$ and $q_1$, 
after $l$ steps the LBD method produces 
\begin{equation}
\begin{aligned}
 &AP_l   =Q_l B_l,& \\
 &A^TQ_l =P_l B_l^T + r_l e_l^T,&\\
 B_l  = &\left( \begin{array}{cccc}
\alpha_1 & \beta_1 &  &  \\
 & \alpha_2 & \ddots  & \\
&  & \ddots & \beta_{l-1}  \\
&  &  & \alpha_l \end{array} \right) &
   = Q_l^T A P_l,
\end{aligned}
\label{eq:LBD}
\end{equation}
where $r_l$ is the residual vector at the step $l$, 
$e_l$ is the $l$-th orthocanonical vector, 
and $Q_l$ and $P_l$ are orthonormal bases of the Krylov subspaces 
$K_l(AA^T, q_1)$ and $K_l(A^TA, p_1)$ respectively. 
Although the latter Krylov space is the same as the one from Lanczos on $C$,
  when LBD is restarted its singular triplet approximations
  often exhibit slow, irregular convergence when the smallest singular values 
  are clustered.
  To address this problem, harmonic projection \cite{kokiopoulou2004computing, baglama2005augmented}, refined projection \cite{jia2003implicitly}, and their combinations \cite{jia2010refined} have been applied to LBD. 
  However, LBD can neither use the locally optimal restarting technique (which is often denoted as +k) \cite{stathopoulos2007nearlyI}, nor make full use of preconditioning. Both techniques become crucial due to the difficulty of the problem even for medium matrix sizes.

The JDSVD method \cite{hochstenbach2001jacobi} extends the Jacobi-Davidson method for singular value problems by exploiting the two
search spaces of the LBD and the special structure of the augmented matrix $B$.
Let $Q$ and $P$ be the bases of the left and right search spaces.
Similarly to LBD, JDSVD computes a singular triplet $(\theta,c,d)$ of the 
projected matrix $H = Q^T A P$ (note that $H$ is not bidiagonal) 
and yields $(\theta,\tilde u, \tilde v)=(\theta,Qc,Pd)$ as an
approximate singular triplet.
JDSVD expands $Q, P$ with the orthogonalized corrections $s$ and $t$ 
for $\tilde u$ and $\tilde v$, which are obtained by solving the 
correction equation
\begin{equation}
\left( \begin{array}{cc}
\Pi_{\tilde u}& 0 \\
0 & \Pi_{\tilde v}  \end{array} \right)
\left( \begin{array}{cc}
-\theta I_m & A \\
A^T & -\theta I_n  \end{array} \right)
\left( \begin{array}{cc}
\Pi_{\tilde u} & 0 \\
0 & \Pi_{\tilde v}  \end{array} \right)
\left( \begin{array}{c}
s \\
t  \end{array} \right)  =
\left( \begin{array}{c}
A v - \theta u \\
A^T u - \theta v \end{array} \right),
\label{eq:JDSVD_correction}
\end{equation}
where $\Pi_{\tilde v} = I_n - \tilde v\tilde v^T, \Pi_{\tilde u} = I_m-\tilde u\tilde u^T$. The JDSVD method can take advantage of preconditioning when solving (\ref{eq:JDSVD_correction}) and, since it does not rely on $P, Q$ being Krylov, can use more effective restarting techniques. 
Because of the two sided projection, it also overcomes the numerical 
problems of working with $C$. However, an optimized eigensolver on $C$ will
still be faster until the numerical issues arise.
  
The SVDIFP method \cite{liang2014computing} extends the EIGIFP inner-outer method \cite{golub2002inverse}. Given an approximate right singular vector and its singular value $(x_i, \rho_i)$ at the $i$-th step of the outer method, it builds a Krylov space ${\cal V} = K_k(M(C - \rho_i I), x_i)$, where $M$ is a preconditioner for $C$. To avoid the numerical problems of projecting on $C$, SVDIFP computes the smallest singular values of $A{\cal V}$, by using a two sided projection similarly to LBD. Because the method focuses only on the right singular vector, the left singular vectors can be quite inaccurate.

\subsection{Current SVD Software}\label{subsec:svd software}

Given the availability of several SVD algorithms, it is surprising that there is a lack of corresponding good quality software, especially with preconditioning.
Table \ref{tab:svd solver software} summarizes the state-of-the-art
  dedicated SVD software packages.
For each package, we show the methods it implements, its programming language,
its parallel computing and preconditioning capabilities, whether it
can obtain a fully accurate approximation efficiently, 
and the main interfaces to these libraries.
The top four are high performance libraries, while the rest
are MATLAB research codes.

SVDPACK \cite{berry1992large} and PROPACK \cite{larsen1998Lanczos} implement variants of Lanczos or LBD methods. In addition, PROPACK implements an implicitly restarted LBD method. 
Both methods work well for computing a few largest, well separated
  singular values, which in most cases is an easy problem.
The computation of smallest singular triplets is not supported in SVDPACK
and it is not efficient in PROPACK which only implements the Rayleigh-Ritz 
projection \cite{jia2010refined}.
In addition, neither library can use preconditioning or support 
  message passing parallelism, although PROPACK does support 
  shared memory multithreading.
These are severe limitations for large-scale problems that need to run 
  on supercomputers and that often converge too slowly without 
  preconditioning.

SLEPc offers an LBD method with thick restarting \cite{andez2008robust}, which has similar algorithmic limitations to PROPACK for computing smallest singular values. In addition, this particular SVD solver cannot be directly used with preconditioning. However, the SLEPc LBD has an efficient parallel implementation.

Despite accuracy limitations, eigenvalue iterative methods based on $C$ are widely used for computing the largest eigenpairs where the loss of accuracy is limited (see (\ref{eq:value-error})) and even for low accuracy computations of the smallest singular values. For example, two popular packages in machine learning, scikit-learn\footnote{\url{http://scikit-learn.org/stable/modules/generated/sklearn.decomposition.TruncatedSVD.html}} and Spark's library MLib\footnote{\url{https://spark.apache.org/docs/1.2.1/mllib-dimensionality-reduction.html}}, use a wrapper for the popular package ARPACK (implicit restarting Arnoldi method) \cite{lehoucq1998arpack}. Other solvers for standard Hermitian eigenvalue problems can also be used. 
Table \ref{tab:eigenvalue solver software} lists the most widely used eigensolver libraries with high-performance computing implementations.

Our hybrid PHSVDS method can leverage these eigensolver libraries to solve 
the partial SVD problem in full accuracy.
However, to optimize for efficiency and robustness several modifications
  and code additions are required that were not all available in previous
  eigensolvers.
For example, Anasazi features a robust, high performance computing 
  implementation but does not provide the near-optimal eigenmethods that
  are critical for fast convergence.
SLEPc relies on PETSc for basic linear algebra kernels with support for various high-performance standards for shared and distributed memory machines and GPU.
It also provides the appropriate preconditioned eigensolvers 
  \cite{hernandez2005slepc}.
However, these are not tuned to deal with the high accuracy requirements 
  of the first stage of PHSVDS or the need for refined projection methods 
  for the highly interior problem in the second stage.
PRIMME is designed to take advantage of all special properties of the Hermitian
  eigenvalue problem and therefore is a natural candidate that only
  required minor extensions, as described later in this paper.
The goal is to produce a high quality, general purpose SVD package that 
  can solve large-scale problems with high accuracy, using parallelism and
  preconditioning, and with as close to a black-box interface as possible.

We mention that in some applications (for example in data mining) the required 
accuracy is so low or the rank of the matrix is so small that the use of 
the power method \cite{lin2010power} or the block power method 
(see Randomized PCA in \cite{2009arXiv0909.4061H}) is sufficient.
These methods cannot be considered as general purpose SVD software and thus 
they are beyond the scope of this paper. 

\begin{table}[t]
   \caption{Dedicated SVD solver software for computing the partial SVD.
   The first four libraries have high performance implementations.
   The rest are MATLAB research codes.
   M, S, G stand for MPI, SMP and GPU, respectively. Fort, Mat, Py, and R stand for Fortran, Matlab, Python, and R programming languages, respectively. }\label{tab:svd solver software}
   \centering
   \scriptsize
   \begin{tabular}{ccccccc}
      HPC library & Method  & Lang & Parallel & Precon. & Fast Full Acc. & Main Bindings/Ports \\ \hline
      PRIMME & PHSVDS & C & M S  & Y  & Y & Fort, Mat, Py, R \\
      PROPACK  & IRLBD & Fort & S & N & N & Mat \\
      SLEPc & TRLBD/KS & C & M S G & N & N & Fort, Mat, Py \\
      SLEPc & JD/GD+k  & C & M S G & Y & N & Fort, Mat, Py \\
      SVDPACK  & Lanczos & Fort & -- & N & N & -- \\ 
       --  & IRRHLB & Mat & S & N & Y & --\\
       -- & IRLBA\footnotemark & Mat & S & N & Y & Py, R \\
       -- & JDSVD & Mat & S & Y & Y & -- \\
       -- & SVDIFP & Mat & S & Y & Y & --\\
      \hline
   \end{tabular}
\end{table}

\begin{table}[t]
   \caption{Eigenvalue solver software available for computing partial SVD by solving an equivalent Hermitian eigenvalue problems on $B$ or $C$. M, S, G stand for MPI, SMP and GPU, respectively. Fort, Mat, Py, R, and Jul stand for Fortran, Matlab, Python, R and Julia programming languages, respectively.}\label{tab:eigenvalue solver software}
   \centering
   \scriptsize
   \begin{tabular}{clccccl}
     Software & Method  & Lang & Parallel  & Precon. & Main Bindings \\ \hline
      Anasazi & KS/GD/LOBPCG & C++ & M S G & Y & Py \\
      (P)ARPACK & Arnoldi & Fort & M S & N & Mat, Py, R, Jul \\
      BLOPEX   & LOBPCG & C & M S & Y & Mat \\ 
      FEAST    & CIRR & Fort & M S & Y  & -- \\ 
      MAGMA    & LOBPCG & C++ & S G & Y & --\\ 
      PRIMME   & JD(QMR)/GD+k/LOBPCG & C & M S & Y & Fort, Mat, Py, R \\
      Pysparse & JD & Py & S & Y & -- \\
      SciPy    & LOBPCG & Py & S & Y & -- \\
      SLEPc    & KS/JD/GD+k & C & M S G & Y & Fort, Mat, Py\\
      SPRAL    & Block & C & S G & N & Fort\\\hline
   \end{tabular}
\end{table}

\section{PRIMME\_SVDS: A High-Performance Preconditioned SVD Software in PRIMME}
\label{sec:primme_svds: a high-performance preconditioned SVD Software}

\footnotetext{The IRBLA method also has a Python port available at 
\url{https://github.com/bwlewis/irlbpy} and a port to R at 
\url{https://github.com/bwlewis/irlba}. Neither port currently supports the 
computation of smallest singular values. The R port can be used on 
parallel environments.}

Our goal is to provide a high quality, state-of-the-art SVD package 
that enables practitioners to solve a variety of large,
sparse singular value problems with \emph{unprecedented efficiency, robustness,
and accuracy}.
In this section we firstly describe our preconditioned two-stage meta-method
proposed in \cite{Wu2015PHSVD} for effectively and accurately computing both
largest and smallest singular values. Then we illustrate in detail the parallel
implementation of PRIMME\_SVDS as well as various characteristics of our
high-performance software. 

\subsection{Method for Efficient and Accurate Computation}
\label{subsec:phsvds method}

The PHSVDS approach \cite{Wu2015PHSVD} relies on eigensolvers that
work on the equivalent eigenvalue formulations $C$ and $B$, switching
from one to the other to obtain the best performance. The PHSVDS method starts
on $C$ because without preconditioning the convergence in terms of iterations 
is much faster than that on $B$, as commented in 
Section~\ref{subsec:svd method}. Furthermore, the cost per iteration (computation of residual vectors,
orthogonalization, etc.) on $C$ is up to two times cheaper than on $B$
(because of dimension $n$ versus $n+m$).
We refer to the computations on $C$ as the \emph{first stage} of the method. If
further accuracy is required, the method switches to a \emph{second stage} where
the eigensolver is reconfigured to work on $B$, but with initial guesses from 
the first stage. We do not consider eigensolvers such as Lanczos that accept 
only one initial guess. Such methods would have to work for each needed 
singular triplet independently which is usually not as efficient as using all
the good quality initial guesses from the first stage in one search space.


The singular value solver is configured to stop when the residual norm of
the required singular triplets is less than the requested 
tolerance $\|A\|_2 \delta_{\text{user}}$, or
\begin{equation}
   \|\tbr_i\|_2 = \sqrt{\|A\tbv_i - \tsigma_i\tbu_i\|_2^2 + \|A^*\tbu_i - \tsigma_i\tbv_i\|_2^2} < \|A\|_2 \delta_{\text{user}}.
   \label{eq:res}
\end{equation}
Let $(\tlambda^C_i,\tbx^C_i)$ be the eigenpair approximation from the eigensolver 
  on $C$.
Considering
that $\tsigma_i$ will be set as $\sqrt{\tlambda^C_i}$, $\tbv_i$ as $\tbx^C_i$ and
$\tbu_i$ as $A\tbx^C_i\tsigma_i^{-1}$, the above is translated into a 
convergence criterion for the eigensolver on $C$ as 
\[
   \|\tbr_i^C\| = \|C\tbx^C_i-\tlambda^C_i\tbx^C_i\|_2 < \sqrt{ |\tlambda^C_i|\|C\|_2} \delta_{\text{user}}.
\]
The eigensolver returns when 
all requested triplets satisfy the convergence criterion.
However, the eigensolver may reach its
maximum achievable accuracy before the residual norm reduces below the
above convergence tolerance.  Setting this limit properly is a critical 
point in the robustness and efficiency of the method. 
If the tolerance is set below this limit the eigensolver may stagnate. 
If the limit is overestimated, then the number of iterations of the second stage 
will increase, making the whole solver more expensive. 
Selecting this limit depends on the numerical properties of the eigensolver, 
so we discuss it in Section~\ref{subsubsec:changes in primme for primme_svds}. 

In the second stage,
the vectors $[\tbv_i; \tbu_i]$ are set as initial guesses of the eigensolver, as follows from \eqref{eq:B_Initguess}.
The convergence criterion directly checks
\eqref{eq:res} with $\tsigma_i$ set as $|\tlambda^B_i|$ and $\tbv_i$ and
$\tbu_i$ set as the normalized subvectors $\tbx^B_i(1:n)$ and $\tbx^B_i(n+1:n+m)$.
Because this computation requires extra reduction operations, it is only checked after
an eigenvalue residual condition is satisfied,
\begin{equation}
   \|\tbr_i^B\| = \|B\tbx^B_i-\tlambda^B_i\tbx^B_i\|_2 \approx \sqrt{2}\|\tbr_i\| < \sqrt{2}\|B\|_2 \delta_{\text{user}}.
\label{eq:B_secondCriterion}
\end{equation}
The above is derived by assuming that $\|\tbx^B_i(1:n)\|_2\approx\|\tbx^B_i(n+1:n+m)\|_2$, where Fortran notation is used for vector subranges.
If the eigenvector $\tbx^B_i$ corresponds to a singular triplet, then this 
  assumption is satisfied near convergence. However, it is possible that a large discrepancy between the two norms exists, 
  $\|\tbx^B_i(1:n)\|_2 \ll \|\tbx^B_i(n+1:n+m)\|_2$, 
  even when \eqref{eq:B_secondCriterion} is satisfied.
This is the case when $\tbx^B_i$ has large components in the 
  $(m-n)$-dimensional null space of $B$, and therefore it does not correspond
  to a singular triplet.
Checking our second level criterion \eqref{eq:res} avoids this problem.
We have observed this situation when computing the smallest singular values 
in problems with condition number larger than $10^8$.

If the smallest singular values are wanted, the eigensolver on $B$ must
compute interior eigenvalues. 
Then, it is important that we specify where the eigensolver should look 
  for them.
First, we want only the non-negative eigenvalues of $B$.
Second, if $m\neq n$ and we ask for small eigenvalues that are very
  close to zero, the eigensolver will keep trying to converge on 
  the unwanted null space.
Therefore, we should only try to find eigenvalues on the right 
  of some positive number.
Third, because this number should be a lower bound to the singular value we seek,
  we can use the perturbation bounds of the approximations of the first stage.
Specifically, we know that 
$\sigma_i \in [\tsigma_i-\|\tbr^C_i\|\tsigma_i^{-1}\sqrt{2}, \tsigma_i]$,
where $\tsigma_i = (\tlambda^C_i)^\frac{1}{2}$ \cite{Wu2015PHSVD}.
Because the augmented approach cannot distinguish eigenvalues smaller 
  than $\|A\|\epsm$,
we configure the eigensolver to find the smallest eigenvalue $\sigma_i$ 
  that is greater than
$\max(\tsigma_i-\|\tbr^C_i\|\tsigma_i^{-1}\sqrt{2}, \|A\|\epsm)$.
This heuristic is also used in \cite{hochstenbach2004harmonic}. 

For interior eigenproblems, alternatives to the Rayleigh-Ritz extraction are recommended, such as the harmonic or the refined variants. As described in the next section, we employ a variant of the refined extraction. 
For each eigenvalue, the shift for the refined extraction is the corresponding
  singular value lower bound.
In PRIMME, these lower bounds are also used as shifts in the Jacobi-Davidson
  correction equation. If the shifts are close enough to the exact eigenvalues, they accelerate the convergence of Jacobi-Davidson.

Algorithm \ref{alg:primme_svds without preconditioning} shows the specific
functionality needed for the two stages of PHSVDS.

\begin{algorithm}
\caption{PHSVDS: a preconditioned hybrid two-stage method for SVD}
\begin{algorithmic}[1]
    \STATEx {\bf Input:} matrix-vector products $A\,\bx$ and $A^T\,\bx$, preconditioner function, global summation reduction, number of singular triplets seeking $k$, tolerance $\delta_{\text{user}}$
    \STATEx {\bf Output:} Converged desired singular triplets $ \{ \tsigma_{i}, \tbu_{i}, \tbv_{i} \}, \ i = 1, \ldots, k $
    
    \item[]
    \STATEx {\bf \textit {First-stage on $C$:}}
    \STATE Set eigensolver matrix-vector as $C=A^TA$ or $AA^T$
    \STATE Set eigensolver convergence criterion: $\|\tbr_i^C\|_2 \leq \max ( \sqrt { |\tlambda^C_i|\|C\|_2 } \delta_{\text{user}}, \epsm\|C\|_2 ) $
    \STATE Run eigensolver seeking largest/smallest eigenvalues of $C$
    \STATE Perform Rayleigh-Ritz on the returned vector basis
    \STATE Set $\tsigma_i=|\tlambda^C_i|^\frac 1 2$, $\tbv_i=\tbx^C_i$ and $\tbu_i=A\tbv_i\tsigma_i^{-1}$
    \IF{all triplets converged with tolerance $\|A\|_2\delta_{\text{user}}$}
       \STATE Return $ \{ \tsigma_{i}, \tbu_{i}, \tbv_{i} \}$, for $i = 1, \ldots, k $
   \ENDIF
   
   \item[]
    \STATEx {\bf \textit {Second-stage on $B$:}} 

    \STATE Set eigensolver initial guesses as 
	$\frac{1}{\sqrt{2}} \left[\begin{array}{c}
	\tbv_{i}\\\tbu_{i}\end{array}\right],\ i = 1, \ldots, k$
   \IF{finding the largest singular values}
        \STATE Set eigensolver extraction method as \emph{standard Rayleigh-Ritz}
        \STATE Set eigensolver to find the largest algebraic eigenvalues
   \ELSE
        \STATE Set eigensolver extraction method as \emph{simplified refined projection}
        \STATE Set eigensolver to find the eigenvalues closest to but greater 
               than\hfill 
               \break \mbox{} \hfill 
               $\max(\tsigma_i-\|\tbr^C_i\|_2\tsigma_i^{-1}\sqrt{2}, \|A\|_2\epsm)$. \hfill\ 

   \ENDIF
   \STATE Set eigensolver convergence criterion as $\|\tbr_i^B\|_2 \leq \sqrt{2}\|B\|_2 \delta_{\text{user}}$, and when it passes check \eqref{eq:res}
   \STATE Run eigensolver on $B$
   \STATE Set $\tsigma_i=|\tlambda^B_i|$, $\tbx^B_i=[\tbv_i; \tbu_i]$, normalize $\tbu_i$ and $\tbv_i$
   \STATE Return $ \{ \tsigma_{i}, \tbu_{i}, \tbv_{i} \}$, for  $i = 1, \ldots, k $
\end{algorithmic}
\label{alg:primme_svds without preconditioning}
\end{algorithm}

\subsection{Descriptions of changes in PRIMME} \label{subsubsec:changes in
primme for primme_svds}

To support PRIMME\_SVDS, we have implemented many enhancements to PRIMME,
including a user defined convergence criterion,
improved numerical quality of converged eigenvectors,
improved robustness to achieve convergence near machine precision, 
a simplified refined projection method, a different locking
scheme for interior eigenvalues, a new scheme for initializing the search space,
and finally a new two-stage meta-method interface. 

To achieve the required accuracy at the first stage, we have to 
  adjust the convergence tolerance at every step based on the value 
  of the current eigenvalue. 
This was not possible in the original PRIMME implementation in
  which $\delta_{\text{user}}$ was set as a user input parameter.
In the new version, we have added a function pointer into PRIMME main 
data structure to allow the user to provide their own convergence 
 test function.
Our top level SVD interface provides line 2 of 
  Algorithm \ref{alg:primme_svds without preconditioning} as the 
  default convergence test function.
  
When PRIMME uses soft locking (i.e., no locking), before exiting it now 
performs an additional Rayleigh-Ritz on the converged Ritz vectors 
to adjust the angles of the desired Ritz vectors. The resulting Ritz 
vectors have improved quality that has proved helpful in the 
second stage \cite{Wu2015PHSVD}. This is mentioned
in Line 4 of the algorithm.

As an iterative method based on matrix-vector multiplications by $C$, the
maximum accuracy that the eigensolver can obtain should be close to
$\|C\|_2\epsm$. We observed that in slow converging cases PRIMME 
eigensolvers may stagnate when the residual norm is still 10 or 
100 times above that limit.
A brief analysis reveals that the major propagation of error occurs
  when restarting the search space $V$ and the auxiliary matrix $W=AV$ as $Vy$ and $Wy$ respectively
  \footnote{As in most Davidson-type methods, PRIMME stores $W$ to allow the computation of the residual without an additional matrix-vector operation. Note also that in this section $V$ refers to
the search space, not the exact singular vectors.}.
Despite $\|V\|_2 =1=\|y\|_2$, the operation occurs a number of times 
  equal to the number of restarts, and thus the expected accumulated 
  error increases by a factor of $\sqrt{\text{restarts}}$.
This factor is more problematic for $W$ where the accumulated
  error becomes $\sqrt{\text{restarts}}\|C\|_2\epsm$, thus 
  preventing the residual to converge to full accuracy.
This was also confirmed experimentally.
Our solution was to reset both matrices, by fully reorthogonalizing 
  $V$ and computing $W=AV$ directly,
  when $\|\tbr_i^C\| < \sqrt{\text{restarts}}\|C\|_2\epsm$, 
  where restarts is the number of restarts since last resetting.
This change has returned the stagnation level to less than
$10\|C\|_2\epsm$ facilitating a very accurate solution at the first stage.

To address the interior eigenproblem of the second stage, we have implemented
  a refined extraction procedure in PRIMME.
The refined procedure computes an eigenvector approximation $\tbx_i$ in the 
  span of $V$ that minimizes the norm of the residual 
  $\|(B-\tau I)\tbx_i\|/\|\tbx_i\|$.
In general, $\tau$ should be as close as possible to the eigenvalue so most
  implementations set it as the Ritz or harmonic Ritz value from the current 
  search space at every iteration 
  \cite{jia1997refined,hochstenbach2004harmonic,morgan1991computing}.
The minimization requires the QR factorization of the tall skinny matrix 
  $BV -\tau V$, for a step cost of $O((m+n)g^2)$ flops and $O(g)$ global 
  reductions per iteration, where $g$ is the number of columns of $V$.
This, however, would be too expensive.
In our case, the $\tsigma_i$ from the first stage are very good eigenvalue
  approximations (if $\kappa_i < 10^8$ in \eqref{eq:value-error})
  so there is little gain to updating the shift at every iteration.
This leads to a simplified and much more efficient implementation of the
  refined procedure.
With constant $\tau$, the cost of updating the QR factorization at every 
  iteration is $O((m+n)g)$ and requires only a constant number of synchronizations, the same as the cost of orthogonalization.
Also, $Q$ and $R$ can be restarted without communications;
  if $V$ is restarted as $VY$, then we compute the QR factorization 
  $RY = \tilde Q\tilde R$ and restart $Q$ as $Q\tilde Q$ and $R$ as $\tilde R$.
The factorization of $RY$ involves a matrix of small dimension and can be 
  replicated in every process.

In the second stage we force the PRIMME eigensolver to use locking. 
The earlier version of PRIMME locked converged eigenvectors only at restart, 
  allowing them to improve for a few more steps in the basis.
The new version of PRIMME changes this for interior problems only.
When an eigenvector converges we force a restart and lock it out.
This improves robustness with the Rayleigh-Ritz method since converged 
  interior eigenvalues may become unconverged causing the method to misconverge.
The refined extraction avoids this problem but it still benefits from 
  the new change.
When an eigenvector is locked, the QR factorization for the refined extraction 
  is recomputed with the new target shift.

In PRIMME the search space is initialized as a block Krylov subspace
starting from any available initial guesses or random vectors. 
We have extended the library's setup options to allow for finer user control 
  on the initialization step.
Among other options, the user can now deactivate the Krylov subspace.
We have found this to be helpful because of the good quality initial guesses 
  in the second stage.

\subsection{High performance characteristics of PRIMME\_SVDS}
\label{subsubsec:parallel operations in primme_svds}

Our library extension inherits the design philosophy of PRIMME with respect 
to performance. This is summarized below and its effects on performance in
Table \ref{tab:characteristics of primme_svds software}.
\begin{itemize}
\item The user must provide as function pointers the 
  matrix-vector product and, optionally, the preconditioner application.
\item PRIMME's implementation works for both parallel and sequential runs. 
It follows the SPMD parallelization model, so if the user 
  has distributed the matrix by rows onto processes, each process in PRIMME
  will hold the corresponding local rows of the vectors.
Small objects are replicated across processes.
If the code is used in parallel, in addition to parallel implementations 
  of the matrix-vector and preconditioning operators, the user must also 
  provide a subroutine for the global sum reduction.
\item PRIMME relies on third-party BLAS and LAPACK libraries to achieve
  single and multi-threaded performance of operations with dense matrices 
  and vectors within each SPMD process.
\item The required workspace is allocated internally or may be provided by 
  the user as a block of memory.
\end{itemize}

\begin{table}[t]
   \caption{The parallel characteristics of PRIMME\_SVDS operations in PRIMME.
   g is the number of columns of $V$}\label{tab:characteristics of primme_svds software}
   \centering
   \scriptsize
   \begin{tabular}{cccc}
    Operations & Kernels or Libs  & Cost per Iteration  & Scalability \\ \hline
    Dense algebra: MV, MM, & BLAS (e.g., MKL, ESSL, & O((m+n)*g) & Good \\
    Inner Prods, Scale & OpenBlas, ACML)  &  &  \\  \hline
    Sparse algebra: SpMV, & User defined (eg, PETSc, & O(1) calls & Application \\
    SpMM, Preconditioner & Trilinos, HYPRE, librsb) & & dependent \\  \hline
    Global reduction & User defined (e.g.,  & O(1) calls of size O(g) & Machine  \\
                     & MPI\_Allreduce) &  & dependent \\ \hline
   \end{tabular}
\end{table}

Some libraries, such as SLEPc and Anasazi, use an object oriented 
  abstraction for matrices and vectors, thus externalizing the 
  control over the actual memory and the operations on it.
SLEPc is based on the structures of PETSc and Anasazi defines its own structures
with templates.
The goal is to facilitate optimizations on different memory hierarchies 
  and heterogeneous architectures such as accelerators.
However, this design may increase overhead and induce unnecessary memory 
  copies to conform with the given abstraction.

PRIMME handles the vectors directly as a block of memory, which 
  may allow for fewer memory copies when the matrix-vector product and preconditioning 
  operators receive and return the corresponding local part of the vectors,
  but also in other places in the code.
Moreover, this memory layout allows us to group our most computationally intensive 
  operations (such as the computation of the residuals and restarting of the basis) 
  in special kernels that minimize the memory accesses and thus display better 
  locality for cache performance.
These new kernels are a new addition to the PRIMME library.
Then, all numerical operations are performed through calls to optimized 
  BLAS and LAPACK which are compatible with this memory model.
PRIMME tries to use the highest level BLAS when this is beneficial, 
  e.g., the use of level 3 BLAS when the block size is greater than one.
The only disadvantage of this approach is that calls to accelerator (GPU) enhanced 
  BLAS/LAPACK libraries have to transfer their arguments during every call. 
Future releases will allow the code to work directly on accelerator memory
  communicating to CPU only the small sequential tasks.

Following the SPMD (single program, multiple data) model for parallel programming, only the largest data 
  structures in PRIMME are distributed; specifically the eigenvectors 
  to be returned, the vectors in the search space $V$, the auxiliary 
  vectors $W=AV$, and, when refined extraction is used, the array $Q$ 
  that holds the orthogonal matrix of the QR factorization of $BV-\tau V$.
The cost of small matrix operations (such as solving the small 
projected eigenvalue problem) is negligible and the operation is duplicated 
  across all processes.
Long vector updates are performed locally with no communication.
Inner products involve a global sum reduction which is the only 
communication primitive required and is provided by the user. 
To reduce latency, PRIMME blocks as many reductions together as possible 
  without compromising numerical stability.

Most, but not all, applications use PRIMME with the MPI framework that 
  nowadays can be used effectively even on shared memory machines.
This avoids the need to store the matrix or the preconditioner on each core. 
Similarly, pure shared memory parallelism or optimized sequential execution
  is obtained by simply linking to the appropriate libraries.

The user-provided matrix-vector and preconditioning operators must be 
  able to perform operations with both the matrix and its transpose 
  and implement the desired parallel distribution in each case.
Note that the parallel behavior of the two stages might be very different
  for rectangular matrices where $n \ll m$.
The PRIMME\_SVDS interface also allows the user to pass functions 
  for performing matrix vector and preconditioning operations with $B$
  and $C$ directly.
This is useful as there are many optimizations that a user can perform 
  to optimize these operations (especially on $B$, see \cite{AS_biJD}).
This feature is not available in other software.

\subsection{Interfaces of PRIMME\_SVDS}
PRIMME\_SVDS is part of PRIMME's distribution with a native C interface.
We provide several driver programs with calling examples as part of
  the documentation,
  ranging from a simple sequential version with a basic matrix-vector 
  multiplication, to a fully parallel and preconditioned version with 
  such functionality provided by the PETSc library.
In addition, we offer interfaces to Matlab, Python, and R that can be 
  used easily by both ordinary and advanced users to integrate with domain 
  specific codes or simply experiment in an interactive environment. 
These interfaces expose the full functionality of PRIMME\_SVDS which,
  depending on the supporting libraries, can include parallelism and 
  preconditioning.

Next, we describe the most important functionality and features of 
  PRIMME\_SVDS using the C interface.
Other interfaces are wrappers that call the C interface.
The problem parameters and the specific method configuration are set 
in a C structure called \code{primme\_svds\_params}.
The most important parameters are the following.
\begin{itemize}
\item \code{m} and \code{n}: the number of rows and columns of the problem
matrix.
\item \code{matrixMatvec(x, ldx, y, ldy, blockSize, transpose, primme\_svds)}:\break
function pointer to the matrix-vector product with $A$. 
The result of the product is stored in $\by$.
If \code{transpose} is zero,
the function should compute $A\bx$, otherwise $A^*\by$.
$\bx$ and $\by$ are matrices with \code{blockSize} columns and
leading dimensions \code{ldx} and \code{ldy} respectively. 
\code{primme\_svds} is included to provide access to all
   \code{primme\_svds\_params} parameters.
\item \code{matrix}: (optional) pointer to user data for \code{matrixMatvec}.
\item \code{numSVals}: the number of desired singular triplets to find.
\item \code{target}: select which singular values to find: the smallest, the
largest or the closest to some value (not discussed in this paper).
\item \code{eps}: the desired accuracy for wanted singular triplets, 
   see \eqref{eq:res}.
\end{itemize}
To run in parallel SPMD mode, the matrix-vector operator must be parallel, 
every process should have the same values for \code{eps}, 
\code{numProcs}, and \code{numSVals}, and the following parameters must be set.
\begin{itemize}
\item \code{numProcs}: the number of MPI processes (must be greater than 1).
\item \code{procID}: the rank of the local process (e.g., the MPI rank).
\item \code{mLocal} and \code{nLocal}: the number of rows and columns local to 
this process. In the parallel \code{matrixMatvec},
$\bx$ and $\by$ address the corresponding local parts, with
$\bx$ having \code{nLocal} rows and $\by$ having \code{mLocal} rows.
\item \code{globalSumDouble}: function pointer to the global sum reduction.
\end{itemize}
These parallel environment parameters are also required in other frameworks 
  such as PETSc (see Sec 3.1 in PETSc user manual \cite{balay2014petsc}) 
  and Tpetra (see class CsrMatrix \cite{baker2012tpetra}).
In shared memory environments, the user may choose to run a single process 
  and link to a threaded BLAS library such as OpenBLAS \cite{xianyi2012model}.

The following optional parameters may be used to accelerate convergence.
\begin{itemize}
\item \code{applyPreconditioner(x, ldx, y, ldy, blockSize, mode, primme\_svds)}:
function pointer to the preconditioning; the function applies the
preconditioner to $\bx$ and stores it into $\by$. 
\code{mode} indicates which operator is the preconditioner for: $A^*A$
(\code{primme\_svds\_op\_AtA}), $AA^*$ (\code{primme\_svds\_op\_AAt}) or $[\b0\
A^*; A\ \b0]$ (\code{primme\_svds\_op\_augmented}).
\item \code{preconditioner}: pointer to user data for
\code{applyPreconditioner}.
\item \code{maxBasisSize}: the maximum number of columns in the search space
basis.
\item \code{minRestartSize}: the minimum number of columns in the search space
basis for restarting.
\item \code{maxBlockSize}: the maximum number of approximate eigenpairs
to be corrected at every iteration. Larger block size may be helpful for 
multiple or highly clustered singular values and usually improves 
cache and communication performance.
\item \code{primme}: PRIMME parameter structure for first stage.
\item \code{primmeStage2}: PRIMME parameter structure for second stage.
\end{itemize}
The default maximum basis dimension (\code{maxBasisSize}) for the eigensolvers
is 15 and the dimension after restarting (\code{minRestartSize}) is 6 if finding
less than 10 largest singular values. Otherwise restarting parameters are
set to 35 and 14 respectively. The default block size (\code{maxBlockSize}) is 1.

All \code{primme\_svds\_params} parameters can be modified by the user.
Furthermore the user can tune individual eigensolver parameters 
in \code{primme} and \code{primmeStage2} for each stage respectively.
Currently, the default method for the first stage is the DYNAMIC method 
of PRIMME which switches dynamically between GD+k and JDQMR attempting to 
minimize time. The second stage defaults to the JDQMR method.
Users can change the default eigensolver methods by calling,

\begin{lstlisting}[xleftmargin=15pt,xrightmargin=5pt,numbersep=4pt]
primme_svds_set_method(svds_method, method_stage1, method_stage2,
                       primme_svds);
\end{lstlisting}
\code{method\_stage1} and \code{method\_stage2} can be any PRIMME preset method.
With this function the user can also change the PHSVDS to a different SVD method, for example to perform only a single stage with the normal equations
(\code{primme\_svds\_normalequations}) or the augmented approach
(\code{primme\_svds\_augmented}). Future versions may include other 
methods such as LBD and JDSVD.

Other advanced features include the following.
\begin{itemize}
\item \code{initSize}: the number of singular vectors provided as initial
guesses.
\item \code{targetShifts}: contains the values closest to which we should 
  find singular values. Only accessed if \code{target} has been set to find 
  interior singular values.
\item \code{numTargetShifts}: the number of values in \code{targetShifts}.
\item \code{numOrthoConst}: number of singular vectors provided as external
orthogonalization constraint vectors (see explanation below).
\item \code{printLevel}: specifies level for printing out information (0--5).
\item \code{outputFile}: the output file descriptor.
\item \code{stats}: the performance report of this run.
\end{itemize}

After specifying the required and optional fields in the structure, 
  we can call the main function:
\begin{lstlisting}[numberstyle=\tiny,xleftmargin=20pt,xrightmargin=20pt,numbersep=5pt]
primme_svds(svals,svecs,rnorms,primme_svds_params)
\end{lstlisting}
The argument \code{svals} and \code{rnorms} are arrays at least of size
\code{numSvals} to store the computed singular values and the residual norms,
computed as in \eqref{eq:res}. Both arrays are filled by all processes. The
argument \code{svecs} is a dense matrix at least of dimension
$(\text{\code{mLocal}}+\text{\code{nLocal}})\times(\text{\code{numSVals}}+
                  \text{\code{numOrthoConst}})$.
If \code{numOrthoConst} is greater than zero, the code will find left (right) 
singular vectors that are orthogonal to a set of \code{numOrthoConst} left (right) constraint vectors. These constraint vectors are provided along with any initial 
guesses in \code{svecs}.
On input, \code{svecs} holds first the \code{numOrthoConst} left constraint vectors 
followed by the \code{initSize} initial left singular vector guesses, each vector 
of size \code{mLocal}. Following these, it holds the same number of right 
constraint vectors and right initial guesses, each vector of size \code{nLocal}.
On output, \code{svecs} holds the left constraint vectors, the converged left
singular vectors, and their right counterparts in this order.
\code{initSize} is updated with the number of converged singular triplets. 

Figure \ref{fig:c interface} illustrates the most basic form of this interface with 
an example in C that computes the four smallest singular values of a rectangular
matrix. Note that most parameters have appropriate defaults that 
do not have to be set (e.g., \code{numOrthoConst} is zero, \code{maxBlockSize} is 1, etc.).
\begin{figure}[tb]
   \begin{lstlisting}[numbers=left,numberstyle=\tiny,xleftmargin=15pt,xrightmargin=5pt,numbersep=4pt]
#include "primme.h" /* header file for PRIMME_SVDS too */ 
double *svals;  /* Array with the computed singular values */
double *rnorms; /* Array with the computed residual norms */
double *svecs;  /* Array with the computed singular vectors */
     /* i-th left (u) vector starts at svecs[primme_svds.m*i],
        i-th right (v) vector starts at
        svecs[primme_svds.m*primme_svds.initSize + primme_svds.n*i] */
/* Create the PRIMME_SVDS configuration struct */
primme_svds_params primme_svds;

/* Set default values in PRIMME_SVDS configuration struct */
primme_svds_initialize(&primme_svds);

/* Set the function that implements A*x and A^t*x */
primme_svds.matrixMatvec = MatrixMatvecSVD;

/* Set problem parameters */
primme_svds.m = 1000;
primme_svds.n = 100;        /* set problem dimension */
primme_svds.numSvals = 4;   /* Number of singular values wanted */
primme_svds.eps = 1e-12;    /* ||r|| <= eps * ||matrix|| */
primme_svds.target = primme_svds_smallest; /* Seek smallest s.v. */

/* Allocate space for converged Ritz values and residual norms */
svals = (double *)malloc(primme_svds.numSvals*sizeof(double));
svecs = (double *)malloc((primme_svds.n+primme_svds.m)
                                *primme_svds.numSvals*sizeof(double));
rnorms = (double *)malloc(primme_svds.numSvals*sizeof(double));

dprimme_svds(svals, svecs, rnorms, &primme_svds);
   \end{lstlisting}
    \caption{Simple sequential example code that computes the four smallest singular values of a rectangular matrix of dimensions $1000\times 100$ with PRIMME. The matrix-vector multiplication code and some details have been omitted. The full version can be found at \code{exsvds\_dseq.c} under the folder \code{examples}.}\label{fig:c interface}
\end{figure}

Finally, we briefly discuss the MATLAB interface which is a MEX wrapper 
of PRIMME\_SVDS. It is similar to MATLAB's {\tt svds}, 
allowing it to be called by non-expert users but also by experts that 
can adjust over 30 parameters.

\begin{lstlisting}[numberstyle=\tiny,xleftmargin=20pt,xrightmargin=20pt,numbersep=5pt]
primme_svds(A)
primme_svds(A, numSvds)
primme_svds(A, numSvds, target)
primme_svds(A, numSvds, target, opts)
primme_svds(A, numSvds, target, opts, precond)
primme_svds(Afun, M, N,...)
\end{lstlisting}
Like \code{svds}, users only need to provide the matrix $A$ while PRIMME\_SVDS
  sets a list of expert defaults underneath.
Users can tackle more advanced tasks incrementally by specifying more
parameters. For example, they can pass their own matrix-vector and 
preconditioning operations, or they can simply
take advantage of MATLAB's built-in matrix times block-of-vectors operators 
and preconditioners. Interfaces to the other scripting languages are developed 
similarly.

\section{Numerical Experiments}\label{sec:experiments}

We report numerical results in order to demonstrate the diverse 
functionalities of PRIMME\_SVDS and to assess its performance relative 
to other software over a wide range of problems.
We seek a few smallest and a few largest singular triplets of matrices 
  of large size ($10^6$ -- $10^7$), with and without preconditioning, 
  and on both distributed and shared memory parallel computers.
We are not aware of other published results for computing smallest singular 
  triplets of such large size and challenging spectrum.
The matrices come from various applications, including least-squares problems,
  DNA electrophoresis, linear programming, and graph clustering, and many
  of the matrices are rectangular. 
The condition number of the non-singular matrices is bounded by $10^4$.
Therefore, for seeking smallest singular values,  
a tolerance of $10^{-6}$ can be achieved by the first stage 
and a tolerance of $10^{-12}$ requires also the second stage.
Basic information on the matrices is listed in Table \ref{tab:mats-info}.  
The first stage of PHSVDS in PRIMME can use GD+k or JDQMR, but the second is set to JDQMR. 
Therefore we include comparisons with both variants. In practice, the first stage in 
PRIMME can use the DYNAMIC method which, with minimal overhead, identifies the best 
of the two variants.

\begin{table}[t]
\caption{Properties of the test matrices. We report the smallest gap ratio
of the five smallest and of the five largest distinct singular values.
For the smallest this is computed as
$\min_{i=1:5, \sigma_i \neq \sigma_{i+1}}\frac{\sigma_i-\sigma_{i+1}}{\sigma_{i+1}-\sigma_n}$, and for the largest as
$\min_{i=n-4:n, \sigma_i \neq \sigma_{i-1}}\frac{\sigma_i-\sigma_{i-1}}{\sigma_{1}-\sigma_{i-1}}$.
}
\centering
\label{tab:mats-info}
\footnotesize
\begin{tabular}{crrrrcccc}
             &            &           &            &            & \multicolumn{2}{c}{gap ratios}\\
 Matrix      &  rows $m$  & cols $n$  &   nnz($A$) & $\kappa(A)$& largest & smallest\\\hline
cage15       &  5,154,859 & 5,154,859 &99,199,551  & 1.2E+1     & 6E-4  & 1E-3\\
atmosmodl    &  1,489,752 & 1,489,752 & 10,319,760 & 1.1E+3     & 5E-5  & 5E-5\\
Rucci1       &  1,977,885 &   109,900 & 7,791,168  & 6.7E+3     & 3E-3  & 5E-5\\
LargeRegFile &  2,111,154 &   801,374 & 4,944,201  & 1.1E+4     & 1.2   & 3E-7\\
sls          &  1,748,122 &    62,729 & 6,804,304  & 1.3E+3     & 4E-2  & 8E-7\\
cont1\_l     &  1,918,399 & 1,921,596 & 7,031,999  & 2.0E+8     & 6E-6  & 5E-8\\
relat9       & 12,360,060 &   549,336 & 7,791,168  &$\infty$    & 3E-3  & --   \\
delaunay\_n24& 16,777,216 &16,777,216 &50,331,601  &$\infty$    & 2E-3  & --   \\
Laplacian    &   8,000$p$ &  8,000$p$ & 55,760$p$  & --         & --    & --  \\ \hline
\end{tabular}
\end{table}

Our computations are carried out on the NERSC's Cray Edison supercomputer 
and on the SciClone cluster at the College of William and Mary. 
On Edison, each compute node has two Intel ``Ivy Bridge'' processors at 2.4 GHz 
for a total of 12 cores and 64 GB of memory,
interconnected with high-speed Cray Aries in Dragonfly topology. On SciClone,
we use 36 Dell PowerEdge R415 servers with AMD Opteron processors at 3.1 GHz
for a total of 12 cores and 32 GB of memory running the Red Hat Enterprise
Linux operating system, interconnected by an FDR InfiniBand communication
network. The machine precision is $2.2 \times 10^{-16}$ in all environments.

\subsection{Comparison with SLEPc LBD on a distributed memory system}

\begin{figure}[t]
   \centering
   \begin{subfigure}[b]{0.49\textwidth}
      \includegraphics[width=\textwidth]{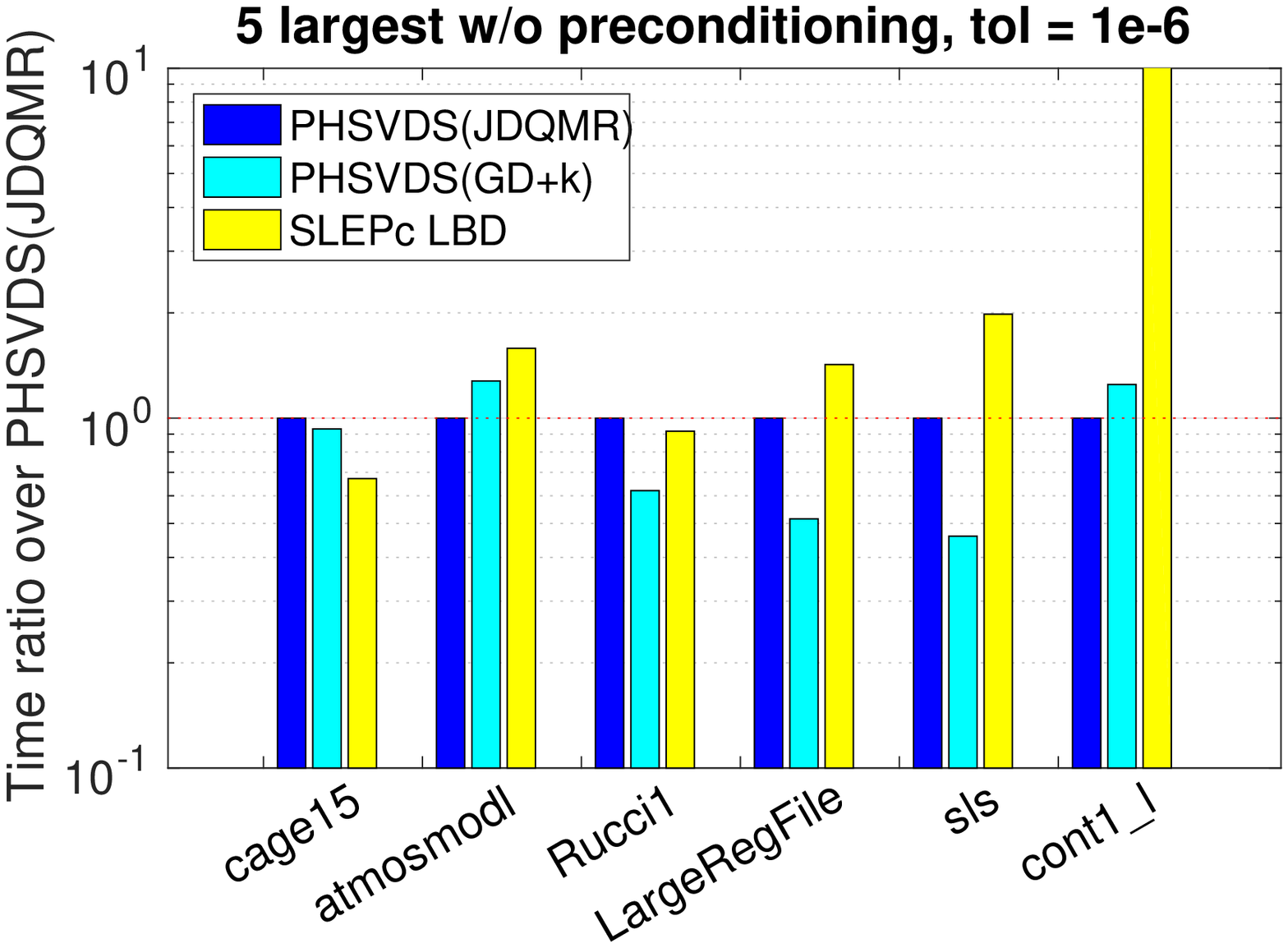}
      \label{fig:dist_comp_la6}
    \end{subfigure}
    \begin{subfigure}[b]{0.49\textwidth}
         \includegraphics[width=\textwidth]{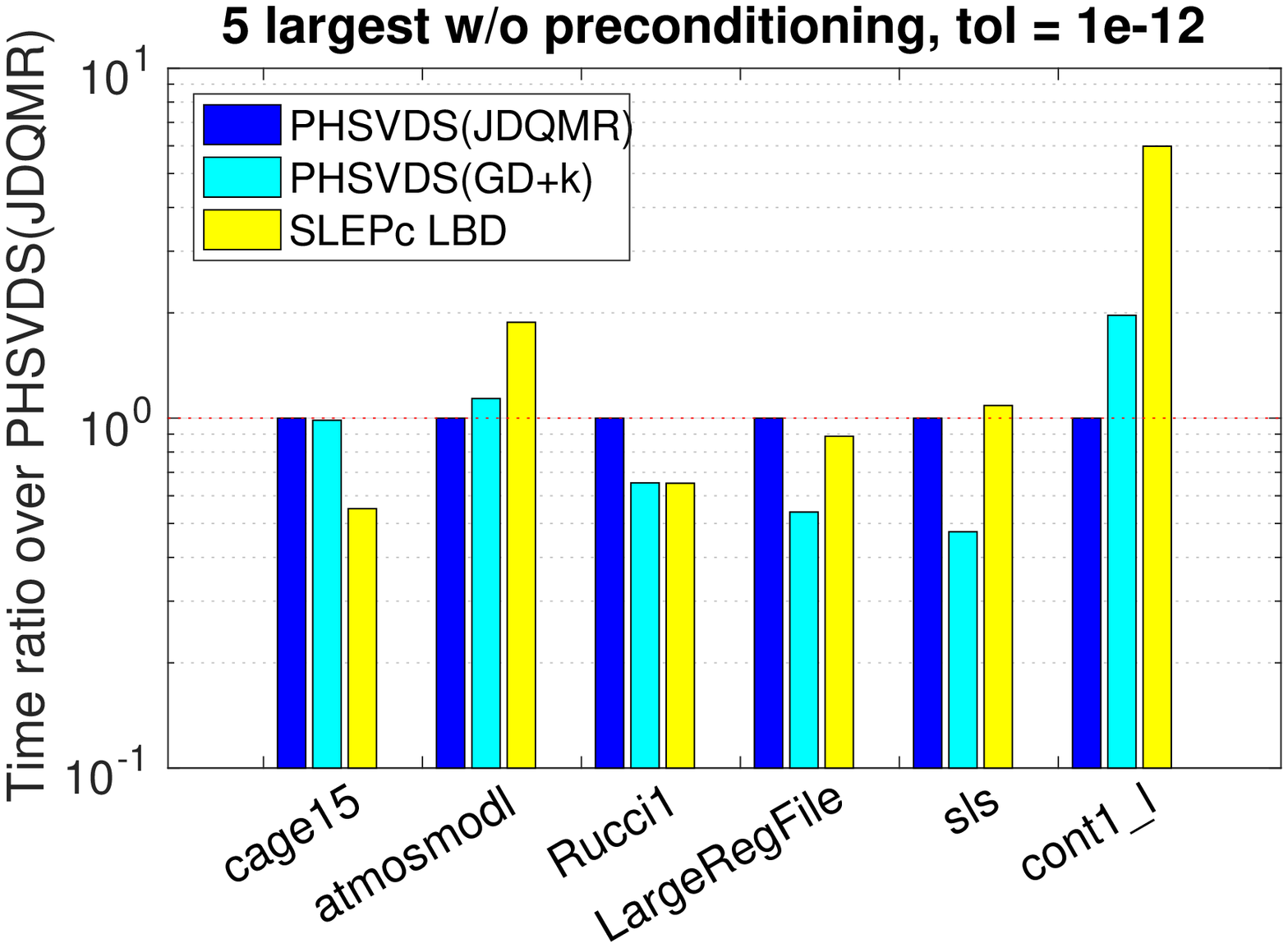}
         \label{fig:dist_comp_la12}
     \end{subfigure}
    \begin{subfigure}[b]{0.49\textwidth}
         \includegraphics[width=\textwidth]{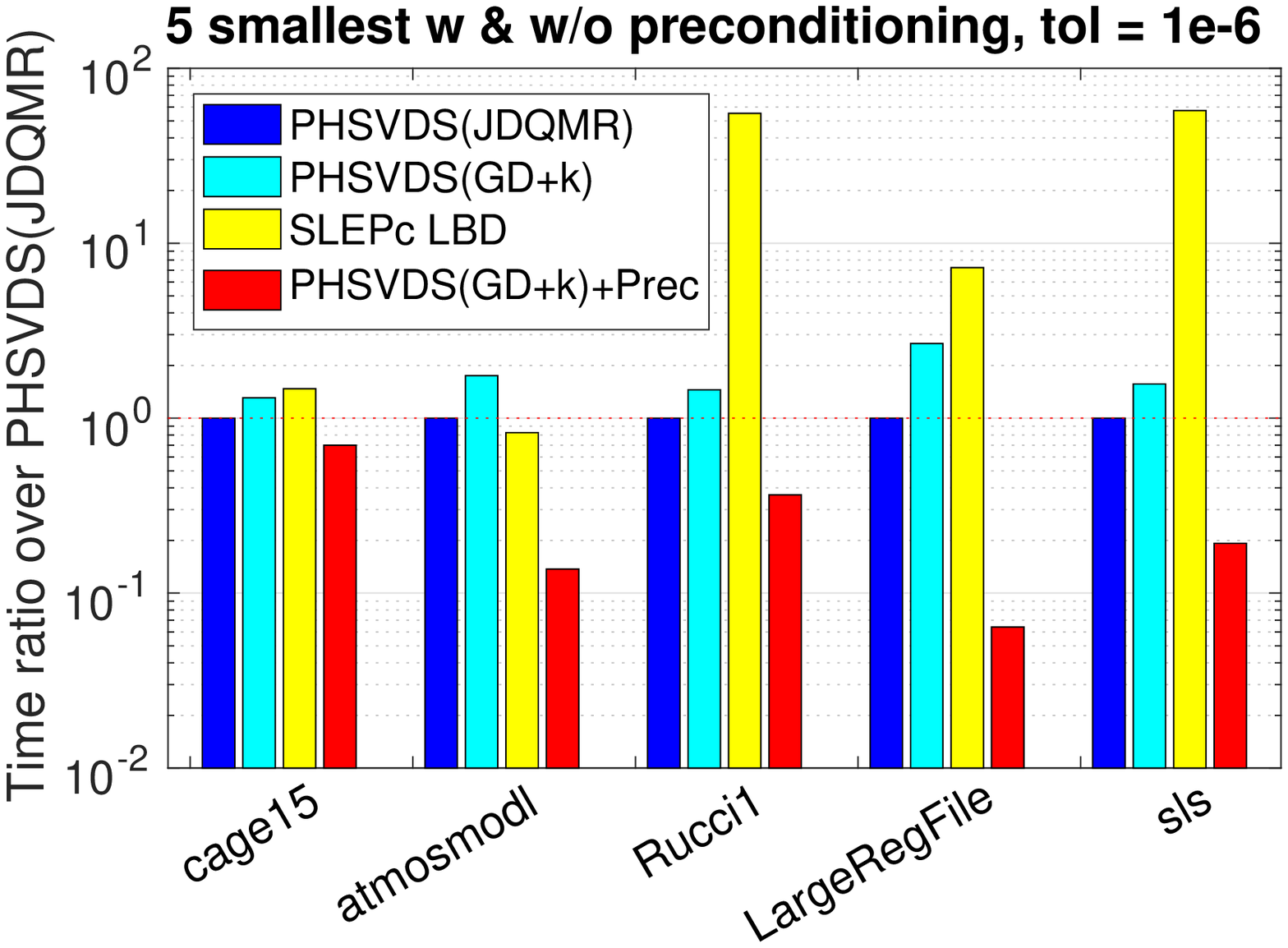}
         \label{fig:dist_comp_sa6}
     \end{subfigure}
    \begin{subfigure}[b]{0.49\textwidth}
         \includegraphics[width=\textwidth]{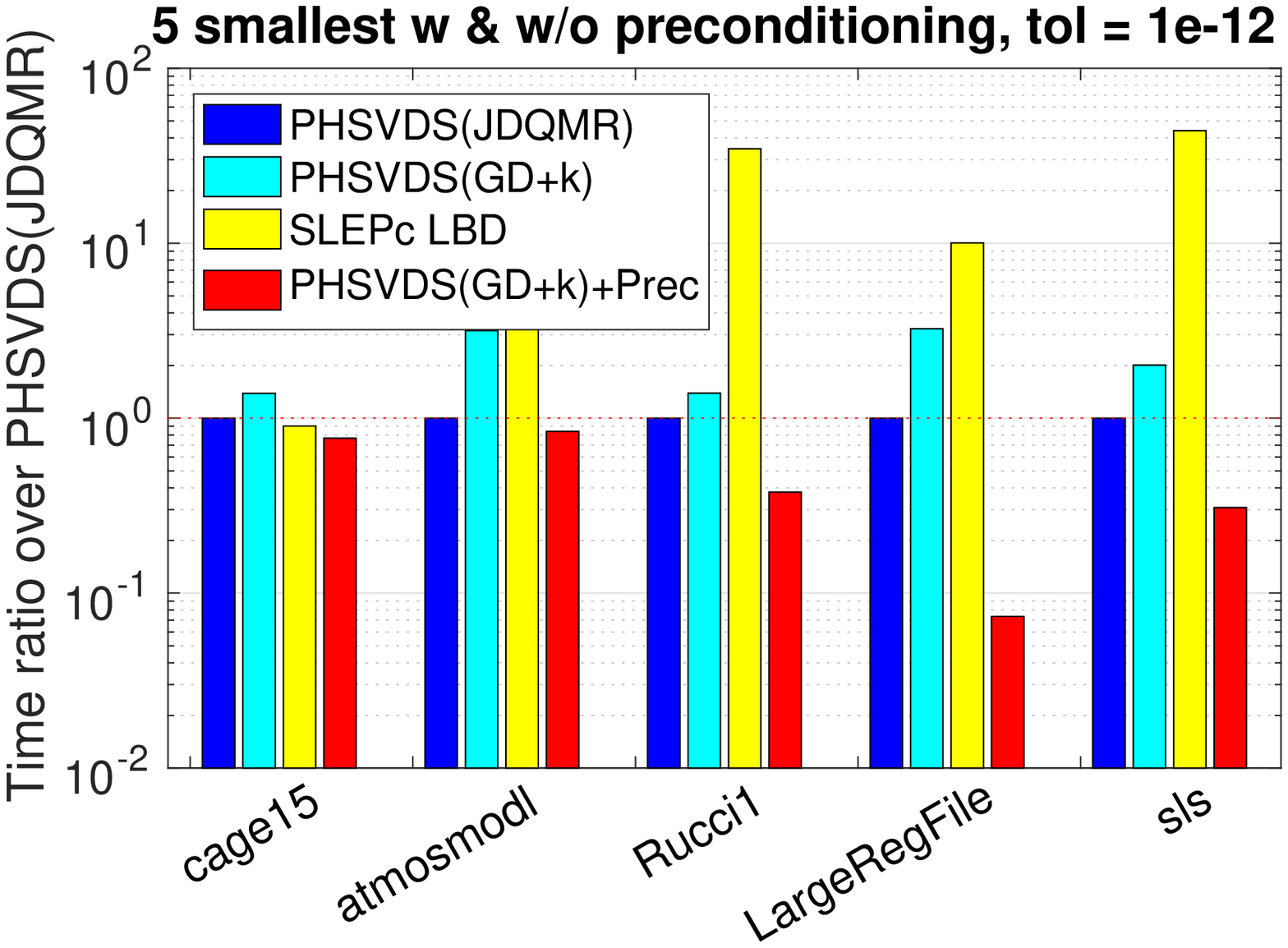}
         \label{fig:dist_comp_sa12}
     \end{subfigure}
     \caption{Time ratio over PHSVDS(JDMQR) when computing 5 largest and
smallest singular values with user tolerance $10^{-6}$ and $10^{-12}$ with
and without
preconditioning using 48 MPI processes in distributed memory. The sparse
matrix-vector operations are performed using PETSc 3.6.0. Note that for atmosmodl,
LBD misses several smallest singular values for $10^{-6}$ and it misses 
  a multiplicity for $10^{-12}$.}
     \label{fig:dist_comp_la_sa_6_12}
\end{figure}

We compare the two PHSVDS variants against thick-restart LBD in 
	SLEPc 3.6.0
  using 48 MPI processes on 4 nodes (i.e., 48 cores) of SciClone.
We compute the 5 largest and the 5 smallest singular values with tolerances
  $10^{-6}$ and $10^{-12}$ in \eqref{eq:res}. 
For the largest singular values, all solvers use a maximum basis size of 15. 
This is the default for PRIMME\_SVDS, but for SLEPc LBD the default 10 
  obtained worse performance in general so it was not used.
For the smallest singular values and when tolerance $10^{-12}$ is needed, 
  PRIMME\_SVDS uses the default basis size of 35 during the second stage.
For the smallest singular values, LBD was run with a basis size of 35 for 
  all tolerances because the basis size of 15 was far slower.
All solvers start with a random initial vector.
The results are summarized in Table~\ref{ta:lbd} and also displayed as time ratios
  between the methods in Figure~\ref{fig:dist_comp_la_sa_6_12}.

\begin{table}[t]
   \caption{Seeking 5 largest and smallest singular triplets, with user
	   tolerance $10^{-6}$ and $10^{-12}$. For $10^{-12}$ we also test preconditioning. 
We report number of matrix-vector operations (two per iteration) and runtime
for PHSVDS(GD+k) and PHSVDS(JDQMR) in PRIMME\_SVDS, and LBD in SLEPc. 
Bold face shows the minimum metric across methods. 
Preconditioning is always faster so its numbers appear in italics.
For atmosmodl, `$^*$' means that the method missed one multiple singular value, 
and `$^{**}$' means that the method missed in addition non-multiple ones.
}
   \label{ta:lbd}
   \small
   \centering
   \newcommand{\rheader}[1]{\multicolumn{2}{c}{#1}}
   \newcommand{\B}[1]{\textbf{#1}}
   \newcommand{\E}[1]{\emph{#1}}
   \newcommand{\Z}{\parbox{0cm}{$^*$}}
   \newcommand{\ZZ}{\parbox{0cm}{$^{**}$}}
   \begin{tabular}{c@{\ \ }r@{\ \ }rr@{\ \ }rr@{\ \ }rr@{\ \ }r}
           & \rheader{P(GD+k)} & \rheader{P(JDQMR)} & \rheader{LBD} & \rheader{Prec} \\
      Matrix         & \multicolumn{1}{c}{MV} & \multicolumn{1}{c}{Sec}&
                       \multicolumn{1}{c}{MV} & \multicolumn{1}{c}{Sec}&
                       \multicolumn{1}{c}{MV} & \multicolumn{1}{c}{Sec}& 
                       \multicolumn{1}{c}{MV} & \multicolumn{1}{c}{Sec}\\\hline\\[-3mm]
      \multicolumn{9}{c}{\emph{the 5 largest singular values with tolerance $10^{-6}$}}\\
      cage15        &      853 &     67.6 &     1165 &     72.6 &  \B{744}&\B{48.8} \\
      atmosmodl     &  \B{1611}&     21.0 &     1991 &  \B{16.4}&    2264 &   26.0 \\
      Rucci1        &      213 &   \B{1.0}&      367 &      1.6 &  \B{184}&    1.5 \\
      LargeRegFile  &  \B{  57}&   \B{0.5}&      129 &      0.9 &      84 &    1.3 \\
      sls           &  \B{  61}&   \B{0.3}&      135 &      0.6 &      78 &    1.3 \\
      cont1\_l      &  \B{1477}&     22.9 &     1931 &  \B{18.4}&   12408 &  201.4\\[1mm]
      \multicolumn{9}{c}{\emph{the 5 largest singular values with tolerance $10^{-12}$}}\\
      cage15        &     2123 &    168.6 &     2751 &    170.9 & \B{1272}&\B{ 94.2} \\
      atmosmodl     &  \B{4983}&     65.3 &     7025 &  \B{57.4}&    8824 &   107.8  \\
      Rucci1        &      469 &  \B{ 2.2}&      763 &      3.3 &  \B{264}&\B{  2.2} \\
      LargeRegFile  &      115 &  \B{ 1.0}&      267 &      1.9 &  \B{ 98}&     1.7  \\
      sls           &      121 &  \B{ 0.6}&      263 &      1.3 &  \B{ 88}&     1.4 \\
      cont1\_l      &    13581 &    215.2 & \B{11953}& \B{109.4}&   35864 &   654.6  \\[1mm]
      \multicolumn{9}{c}{\emph{the 5 smallest singular values with tolerance $10^{-6}$}}\\
      cage15        &  \B{1035}&     81.9 &     1353 &  \B{62.7}&     874 &    92.4   &\E{   317}&\E{  43.9}\\
      atmosmodl     &   63579\Z&   830.3\Z&   58987\Z&   474.4\Z&21104\ZZ &   392.2\ZZ&\E{   195}&\E{  65.1}\\
      Rucci1        &    84767 &    392.0 & \B{84251}& \B{270.2}& 1118532 & 14914.2   &\E{ 11925}&\E{  98.5}\\
      LargeRegFile  &    16613 &    151.3 & \B{11731}& \B{ 56.7}&   15056 &   411.2   &\E{   343}&\E{   3.6}\\
      sls           &    19763 &     97.7 & \B{17333}&  \B{62.3}&  100956 &  3568.6   &\E{  2307}&\E{  12.0}\\[1mm]
      \multicolumn{9}{c}{\emph{the 5 smallest singular values with tolerance $10^{-12}$}}\\
      cage15        &     2309 &    183.2 &     2867 &    132.1 &\B{ 1124} &\B{119.1} &\E{   683}&\E{ 101.4}\\
      atmosmodl     &   260563 &   3394.2 &\B{181489}&\B{1073.0}&  753886\Z& 14176.2\Z&\E{  2751}&\E{ 902.4}\\
      Rucci1        &\B{188395}&    905.0 &   201979 &\B{ 651.0}& 1669040  & 22554.3  &\E{ 26255}&\E{ 246.1}\\
      LargeRegFile  &    48195 &    440.6 &\B{ 27771}&\B{ 135.8}&   49906  &  1363.8  &\E{   867}&\E{  10.0}\\
      sls           &    72317 &    356.7 &\B{ 49659}&\B{ 177.4}&  218134  &  7807.1  &\E{ 10317}&\E{  54.7}\\ \hline
\end{tabular} 
\end{table}

Computing the five largest singular values is a relatively easy problem, as expected
  from Table \ref{tab:mats-info}, and all packages perform similarly well (within less than 
  a factor of two of each other).
It is harder problems that tend to magnify the difference between methods.
For example, PRIMME\_SVDS has a clear advantage when looking for the largest singular 
  values of cont1\_l (10 and 6 times faster than LBD for large and low tolerance,
  respectively).
These differences are far more apparent when looking for the poorly separated
  smallest singular values. 
There, the PRIMME\_SVDS variants significantly outperform the LBD as shown both in 
the table and as time ratios of more than 10 in Figure \ref{fig:dist_comp_la_sa_6_12}.

We also note the increased robustness of PRIMME\_SVDS over LBD.
For cont1\_l, both PRIMME\_SVDS variants identify five singular values smaller 
  than the requested tolerance of $10^{-6}$, and PHSVDS(JDQMR) is the only solver
  that converges to the smallest singular value to tolerance $10^{-12}$ within
  the allowed time. In contrast, LBD fails to converge to a single triplet for
  both tolerances.
The atmosmodl matrix has extremely small gap ratios in the lower part of its
  singular spectrum and a double (multiple) singular value.
When using $10^{-6}$ tolerance, PRIMME\_SVDS finds only one of the two multiple 
  singular values, but LBD misses in addition two other singular values.
For tolerance $10^{-12}$, LBD still misses the required multiplicity while 
  PRIMME\_SVDS variants have no problem identifying it.

Preconditioning is not often considered for SVD problems because of the
  difficulty of preconditioning the matrices $C$ or $B$.
Besides cases where users can exploit a special structure of their problem,
  sometimes generic preconditioners can be used effectively.
We demonstrate this on the above difficult matrices using some black box
  preconditioners without taking into account the structure of the matrices.
The matrices cage15 and atmosmodl are preconditioned with $M^{-1}M^{-T}$, 
  where $M^{-1}$ is a preconditioner of A built with HYPRE BoomerAMG \cite{HENSON2002155}.
For the rest, the preconditioner is based on block Jacobi on $A^TA$ 
  with block size limited to 600. 
The preconditioners are not specifically tuned (we use HYPRE BoomerAMG with 
  the default setup in PETSc and we have not exhaustively tested the impact 
  of the block size in the Jacobi preconditioner).
Nevertheless, we obtain substantial speedups over the best time without 
  preconditioning (in one case by a factor of 50), as shown in Table \ref{ta:lbd}
  and better depicted in Figure \ref{fig:dist_comp_la_sa_6_12}.
PRIMME\_SVDS provides full flexibility so that users can provide preconditioning 
  for $A$ and $A^T$ or directly for $C$ and $B$. 

\begin{figure}[t]
   \centering
   \begin{subfigure}[b]{0.49\textwidth}
      \includegraphics[width=\textwidth]{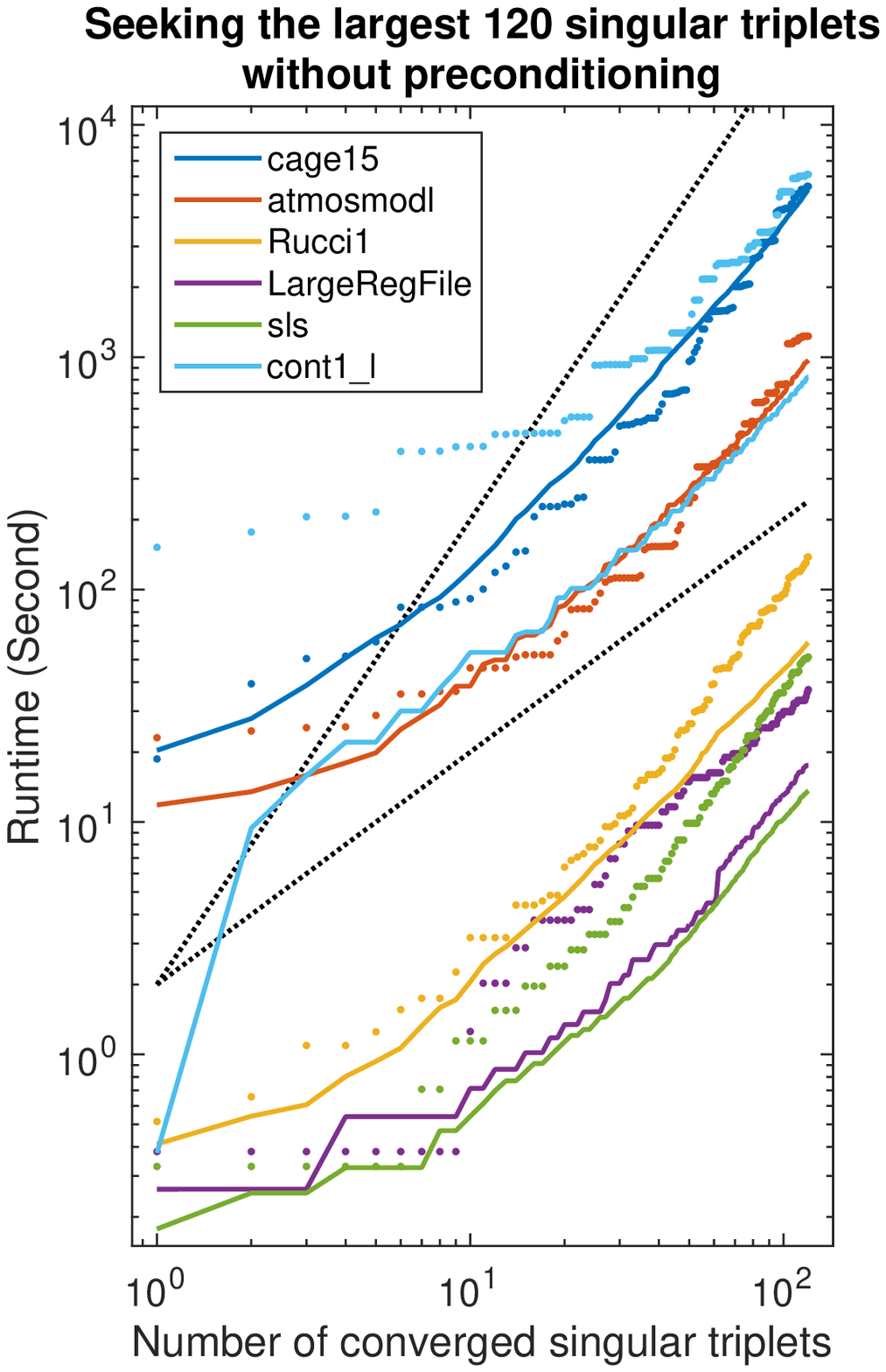}
      \label{fig:la_120}
    \end{subfigure}
    \begin{subfigure}[b]{0.49\textwidth}
      \includegraphics[width=\textwidth]{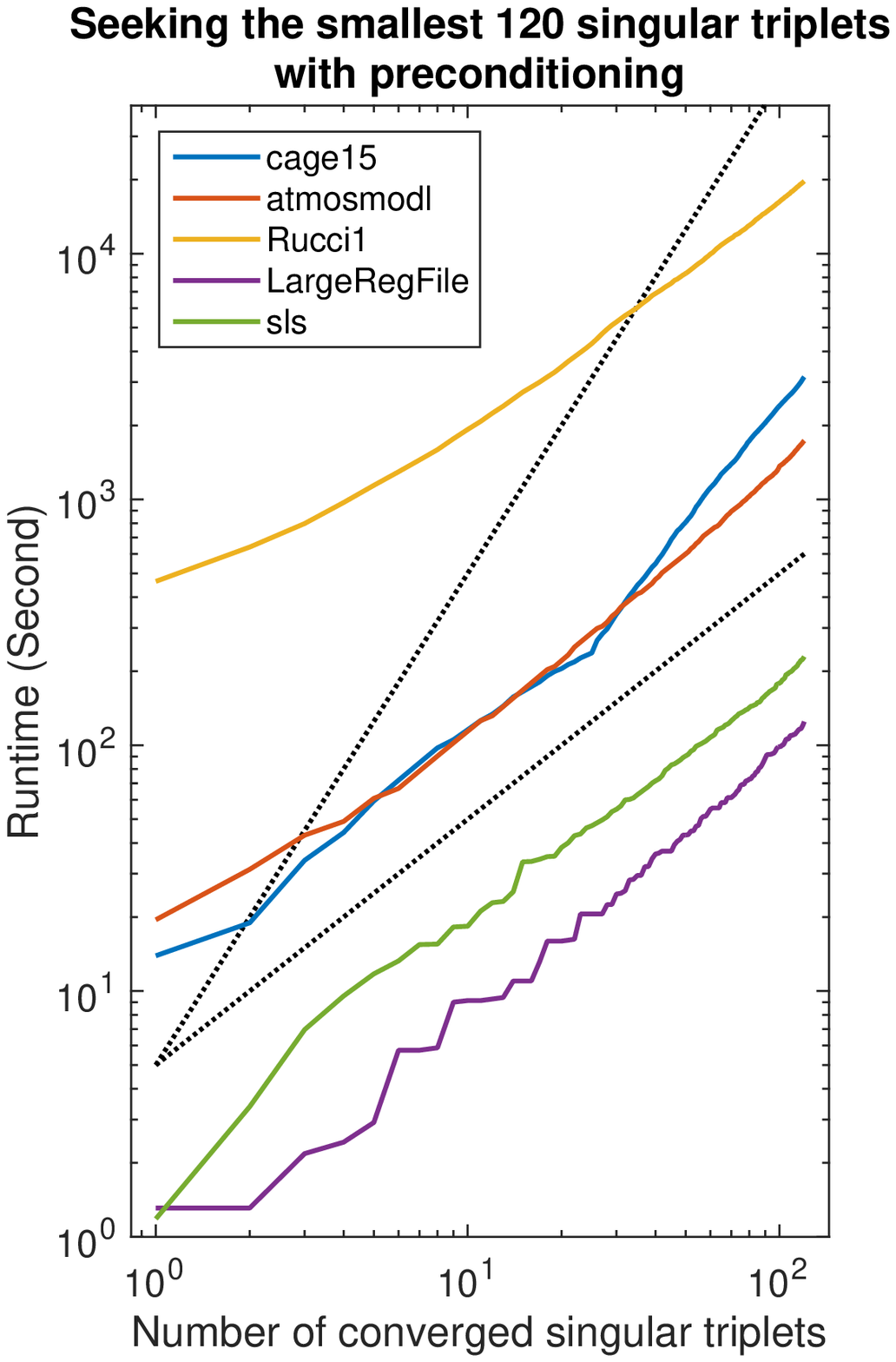}
      \label{fig:sa_120}
     \end{subfigure}
     \caption{Time evolution of computing up to 120 singular values with
tolerance $10^{-6}$ using 48 MPI processes in distributed memory. The sparse
matrix-vector operations are performed using PETSc. Left figure compares
PHSVDS(JDMQR) (colored lines) and SLEPc LBD (colored dots) computing the largest
singular values. Right figure shows PHSVDS(GD+k) computing the smallest singular
values using HYPRE BoomerAMG for atmosmodl and block Jacobi on $A^TA$ with block
size of 600 for the rest. Black dotted lines show reference slopes for linear
and quadratic time costs with respect to the number of singular triplets.}
     \label{fig:la_sa_120}
\end{figure}

Figure \ref{fig:la_sa_120} corroborates that the performance advantages 
of PRIMME\_SVDS extend to the case when we seek a large number of singular 
triplets. 
In the left graph, PHSVDS(JDQMR) is consistently faster than SLEPc LBD 
seeking up to 120 singular triplets from the largest part of the spectrum: 
similar times for cage15 and atmostmodl, 
6 times faster for cont1\_l and 
3 times faster for the rest.
To study the asymptotic scalability with the number of singular triplets,
the figures include two black dotted lines, one for the linear
slope and one for the quadratic. 
As more singular triplets converge, the time for both solvers is 
dominated by the cost of orthogonalization against the locked singular vectors,
approaching the quadratic slope. The use of a preconditioner makes each iteration more expensive but reduces the number of outer iterations, thus
delaying the dominance of orthogonalization.
This is observed in the right graph of Figure \ref{fig:la_sa_120}.

\subsection{Comparison with PROPACK on a shared memory system} 
We compare the performance of PRIMME\_SVDS with PROPACK on a shared 
  memory system.
Both solvers are provided with the multithreaded sparse matrix-vector operation
in librsb \cite{martone2014efficient} and linked with 
a threaded version of BLAS, the AMD Core Math Library (ACML).
We use a single node of SciClone with a total of 12 threads.

Table~\ref{ta:propack} shows the results in terms of matrix-vector products 
  and time.
For PRIMME\_SVDS the results are qualitatively similar to the previous 
  numerical experiments, demonstrating a clear advantage in robustness and 
  performance.
PROPACK, even with fine-tuned settings, has trouble converging to the largest 
  singular values of the most difficult case and to the smallest singular values 
  of almost all cases.
And when it converges it is significantly slower than PRIMME\_SVDS.
It is also slower than the mathematically equivalent LBD method in SLEPc, 
  probably because of the use of partial instead of full reorthogonalization.

\begin{table}[t]
   \caption{Seeking 5 largest and smallest singular triplets with user
tolerance $10^{-6}$ and $10^{-12}$ without preconditioning. We report 
number of matrix-vector operations and runtime from 
PHSVDS(GD+k), PHSVDS(JDQMR) and PROPACK.}
   \label{ta:propack}
   \small
   \centering
   \newcommand{\rheader}[1]{\multicolumn{2}{c}{#1}}
   \newcommand{\B}[1]{\textbf{#1}}
   \newcommand{\E}[1]{\emph{#1}}
   \begin{tabular}{c@{\ \ }r@{\ \ }rr@{\ \ }rr@{\ \ }r}
           & \rheader{P(GD+k)} & \rheader{P(JDQMR)} & \rheader{PROPACK} \\
      Matrix         & \multicolumn{1}{c}{MV} & \multicolumn{1}{c}{Sec}&
                       \multicolumn{1}{c}{MV} & \multicolumn{1}{c}{Sec}&
                       \multicolumn{1}{c}{MV} & \multicolumn{1}{c}{Sec}\\\hline\\[-3mm]
      \multicolumn{7}{c}{\emph{the 5 largest singular values with tolerance $10^{-6}$}}\\
      cage15         &\B{ 872}&  532.9 &    1238 &\B{499.9}&    1640 &  741.7 \\
      atmosmodl      &\B{1824}&  233.3 &    2514 &\B{184.6}&   15308 & 1429.2 \\
      Rucci1         &\B{ 206}&\B{ 8.2}&     426 &    11.2 &     348 &   28.0 \\
      LargeRegFile   &\B{  52}&\B{ 6.6}&     108 &     7.5 &     144 &   24.8 \\
      sls            &\B{  50}&\B{ 3.3}&     154 &     5.4 &     144 &   11.9 \\
      cont1\_l       &\B{1292}&  217.8 &    2990 &\B{210.8}&      -- &     -- \\[1mm]
      \multicolumn{7}{c}{\emph{the 5 smallest singular values with tolerance $10^{-12}$}}\\
      cage15         &\B{ 1054}&    652.7  &    1428 &\B{  600.3} & 1368 &  659.3 \\
      atmosmodl      &\B{64082}&   8603.8  &   69292 &\B{ 5548.4} &   -- &     --\\
      Rucci1         &\B{86072}&\B{2290.5} &  103762 &    2394.1  &   -- &     --\\
      LargeRegFile   &   16464 &   1168.4  &\B{14434}&\B{  530.8} &   -- &     --\\
      sls            &   20134 &    500.9  &\B{18122}&\B{  390.5} &   -- &     -- \\ \hline
\end{tabular} 
\end{table}

\subsection{Strong and Weak Scalability}
We investigate the parallel performance of various methods in PRIMME\_SVDS
and in SLEPc on a distributed memory system. All methods use 
the parallel sparse matrix-vector operation in PETSc.

\begin{figure}[t]
   \centering
   \begin{subfigure}[b]{0.49\textwidth}
      \includegraphics[width=\textwidth]{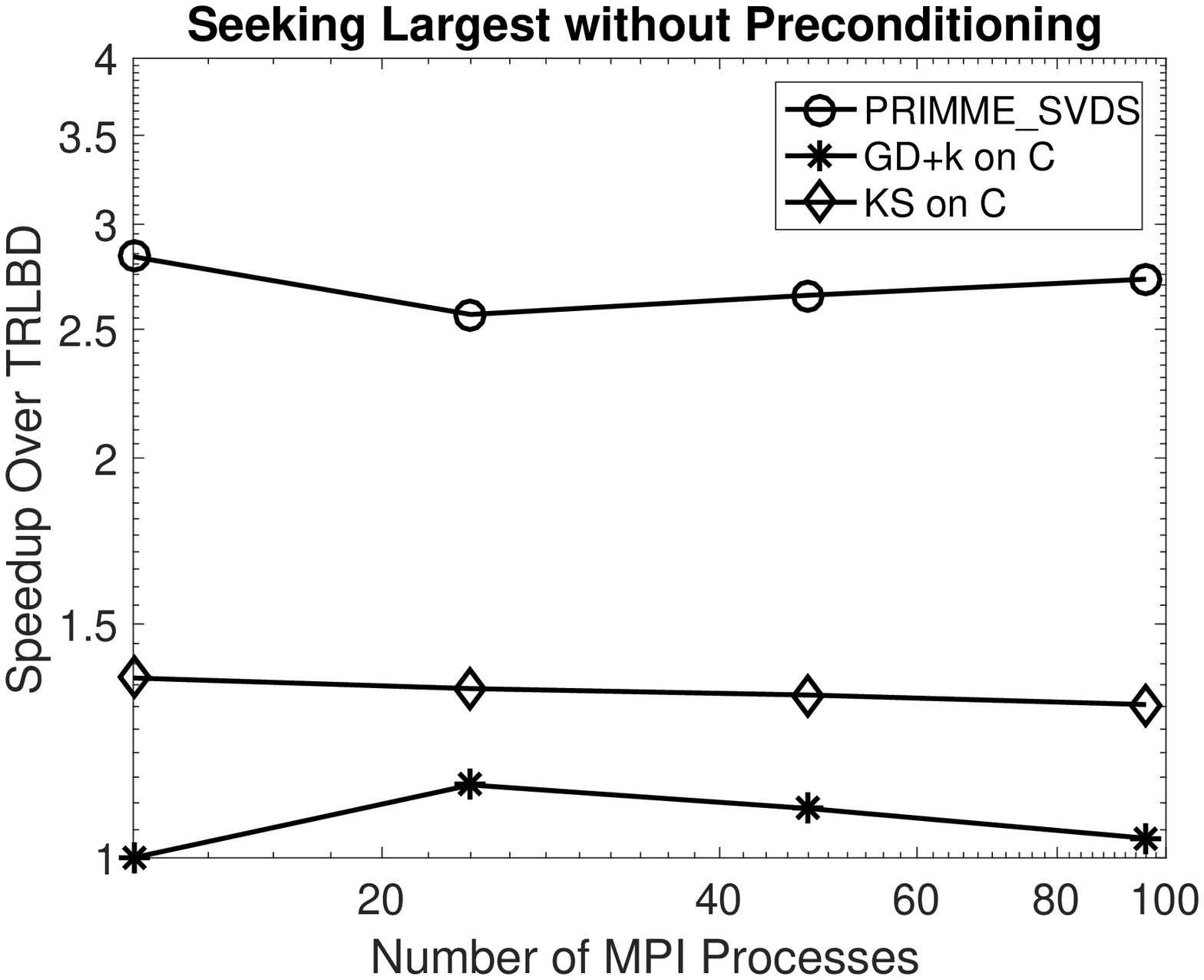}
      \label{fig:dist_comp_la6_varyproc}
    \end{subfigure}
    \begin{subfigure}[b]{0.49\textwidth}
         \includegraphics[width=\textwidth]{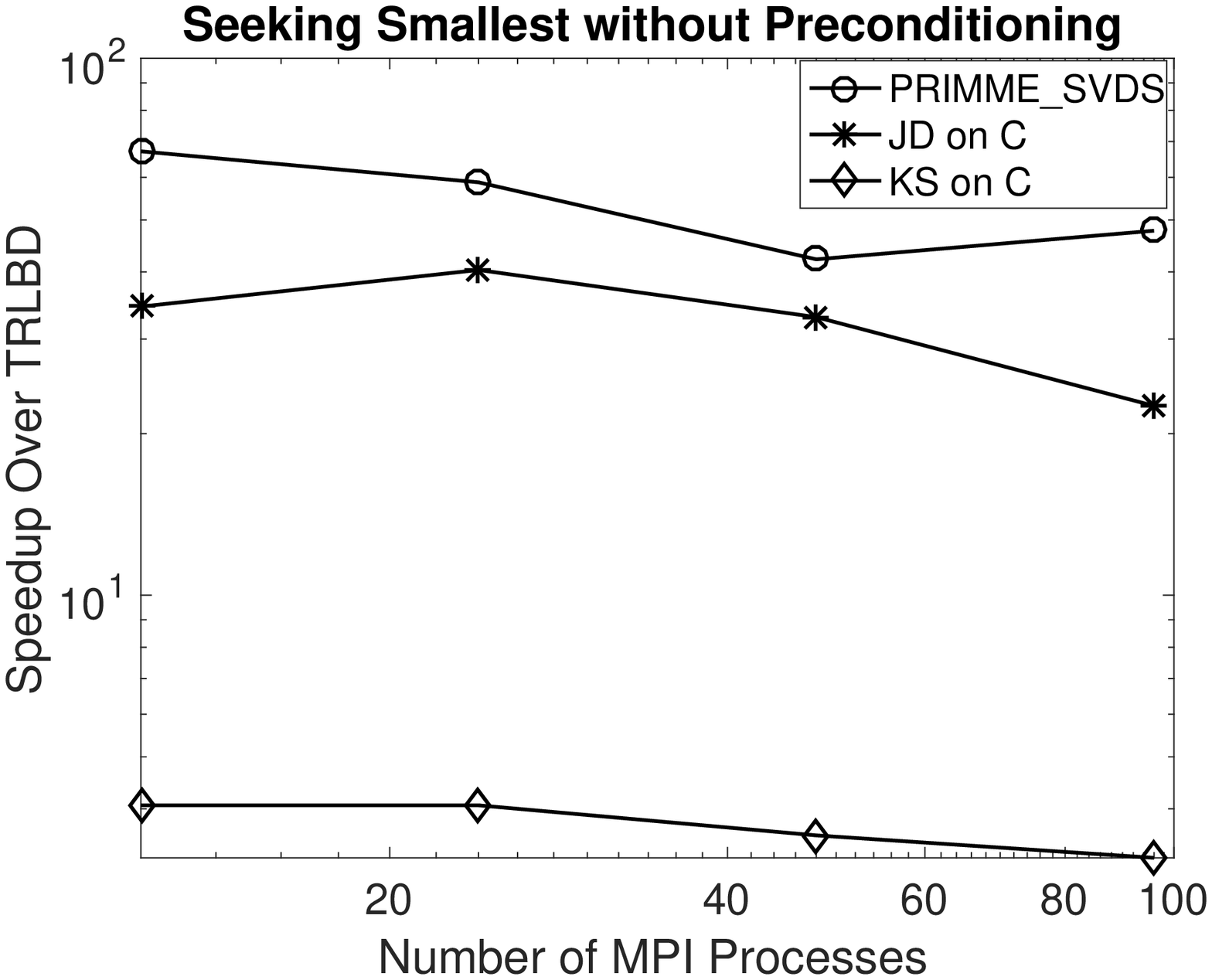}
         \label{fig:dist_comp_sa6_varyproc}
     \end{subfigure}
     \caption{Speedup over SLEPc LBD computing 10 largest and smallest singular
values on delaunay\_n24 (left) and relat9 (right) with user tolerance $10^{-6}$
without preconditioning when increasing the number of MPI processes on SciClone.
The GD+k and KS methods on $C$ are eigensolver implementations in SLEPc.
The sparse matrix-vector operations are performed using PETSc.}
     \label{fig:dist_comp_la_sa_6_varyproc}
\end{figure}

In the first comparison, we report speedups in time of various methods over 
  SLEPc's LBD method as we vary the number of processes from one to 100.
The other methods are the GD+k method in the first stage of PRIMME\_SVDS, 
  and GD+k, Jacobi-Davidson (JD), and Krylov-Schur (KS) as implemented in SLEPc,
  all operating on $C$.
The runs were made on SciClone and the results are shown in 
  Figure \ref{fig:dist_comp_la_sa_6_varyproc}.
While PHSVDS clearly outperforms the rest of the solvers in terms of time, 
  what is more relevant is that the ratio for PHSVDS keeps almost constant 
  with the number of processes. 
This is an indicator that PRIMME\_SVDS has similar parallel scalability
  with LBD, and better scalability than other SLEPc solvers 
  (including the similar GD+k).
Notice that the matrix of the left plot, delaunay\_n24, is extremely sparse
  with only 3 elements per row, so this scalability study reflects better
  the parallel efficiency of the solver and much less of the PETSc matvec.


\begin{figure}
   \centering
   \begin{subfigure}[b]{0.49\textwidth}
      \includegraphics[width=\textwidth]{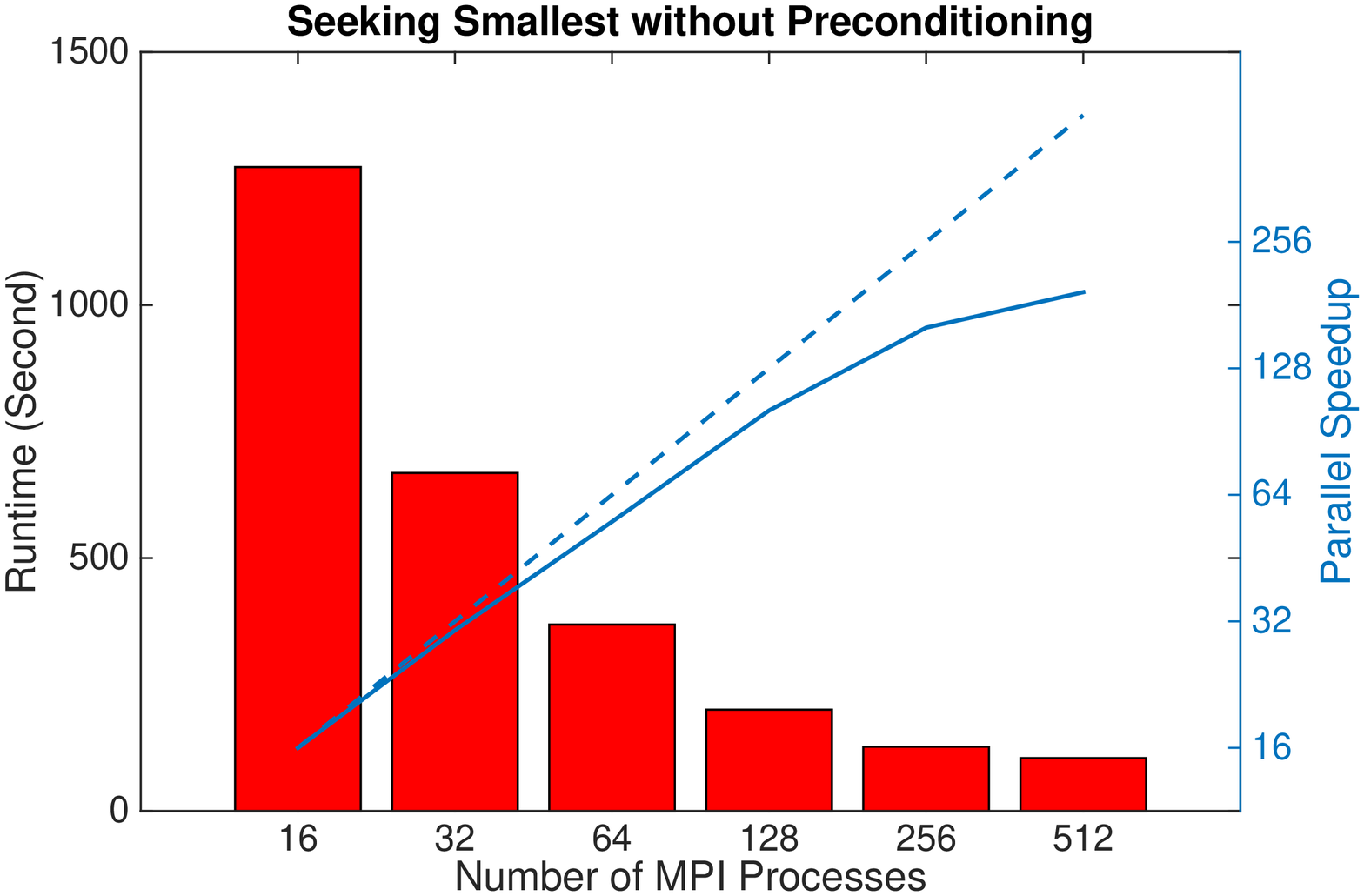}
      \caption{}
      \label{fig:StrongScaling_rectangular_SA}
    \end{subfigure}
    \begin{subfigure}[b]{0.49\textwidth}
         \includegraphics[width=\textwidth]{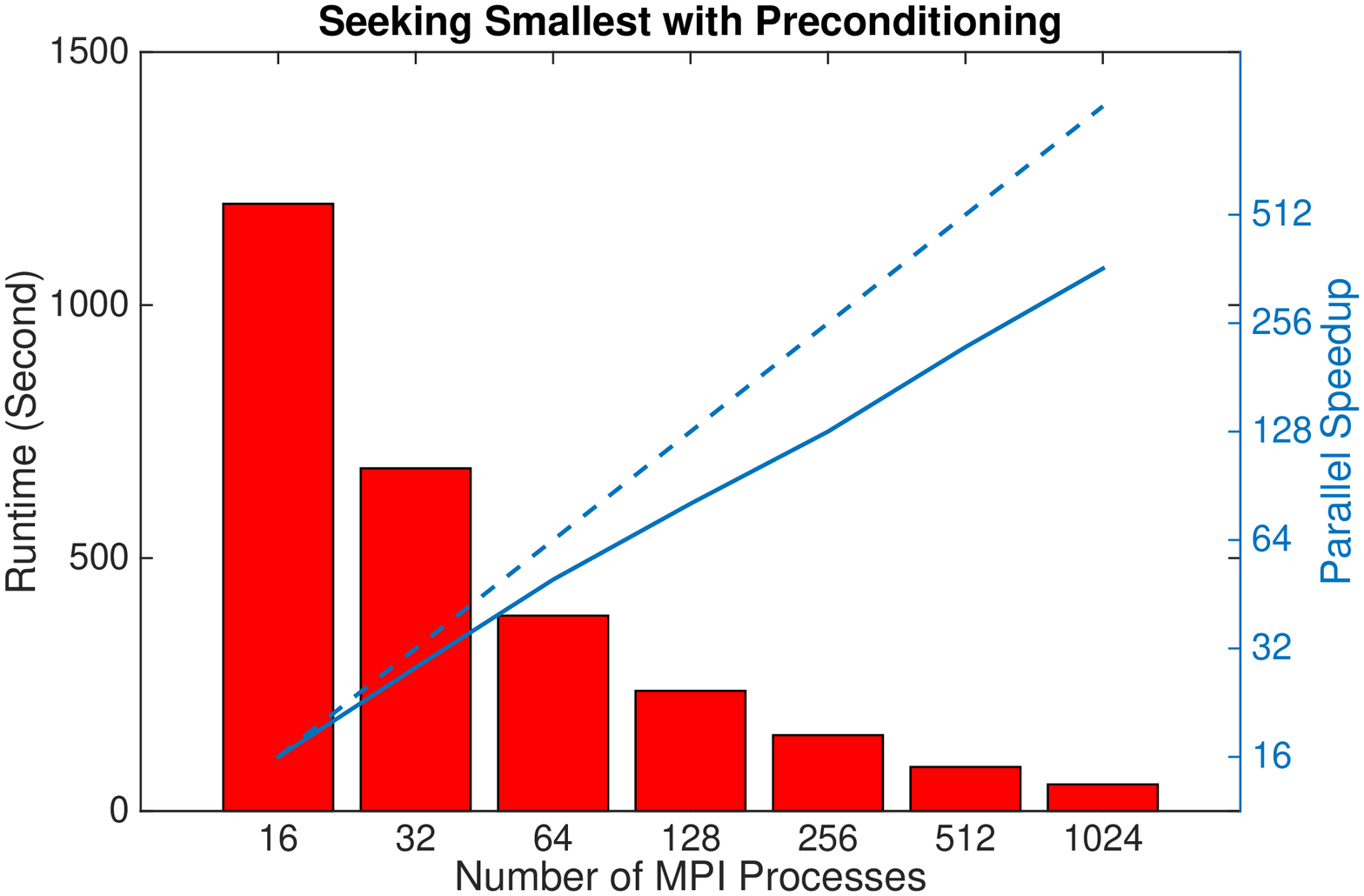}
         \caption{}
         \label{fig:StrongScaling_square_SA_Prec}
     \end{subfigure}
     \label{fig:StrongScaling_rectangular}
   \begin{subfigure}[b]{0.49\textwidth}
      \includegraphics[width=\textwidth]{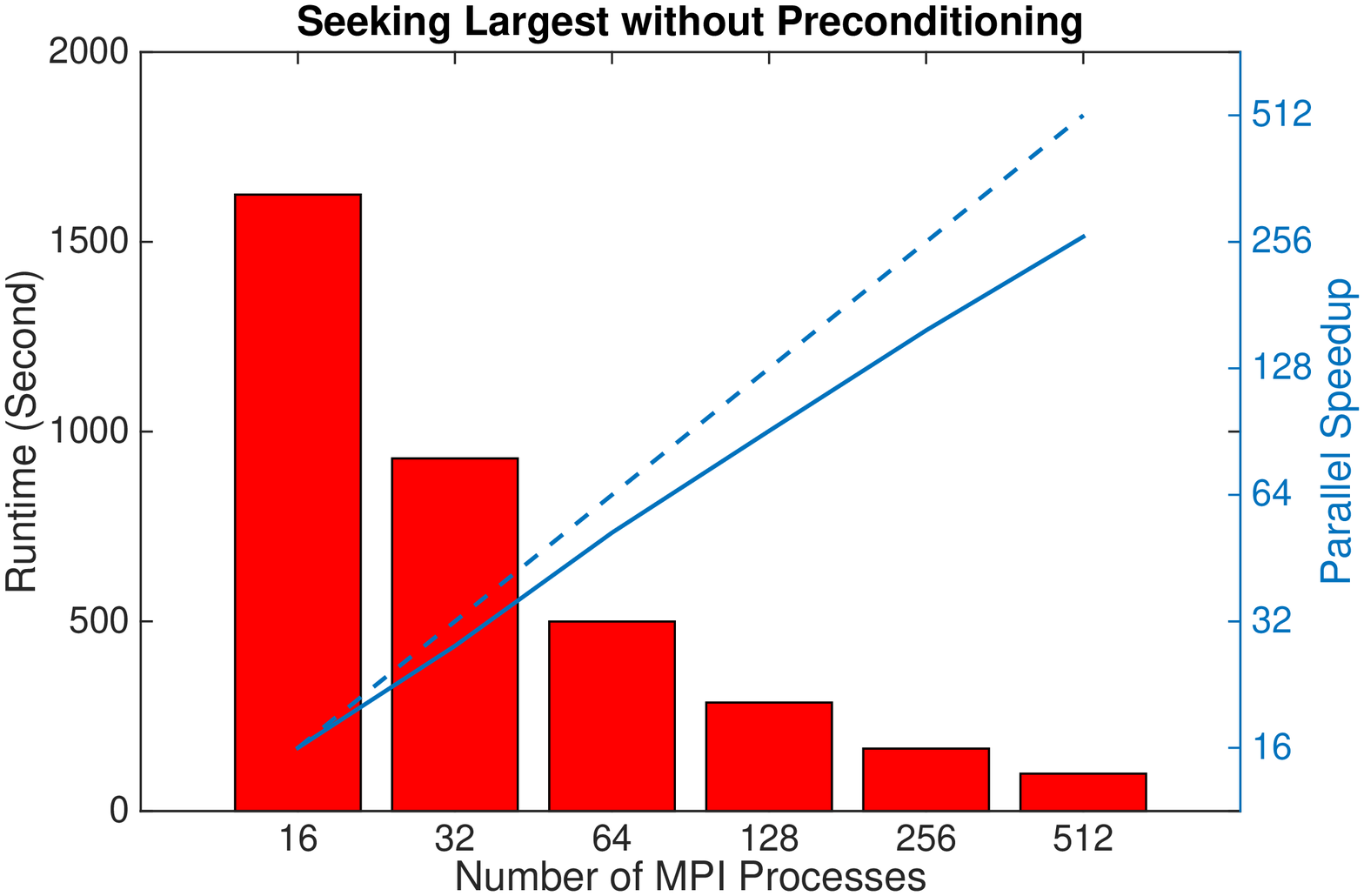}
      \caption{}
      \label{fig:StrongScaling_square_LA}
    \end{subfigure}
    \begin{subfigure}[b]{0.49\textwidth}
      \includegraphics[width=\textwidth]{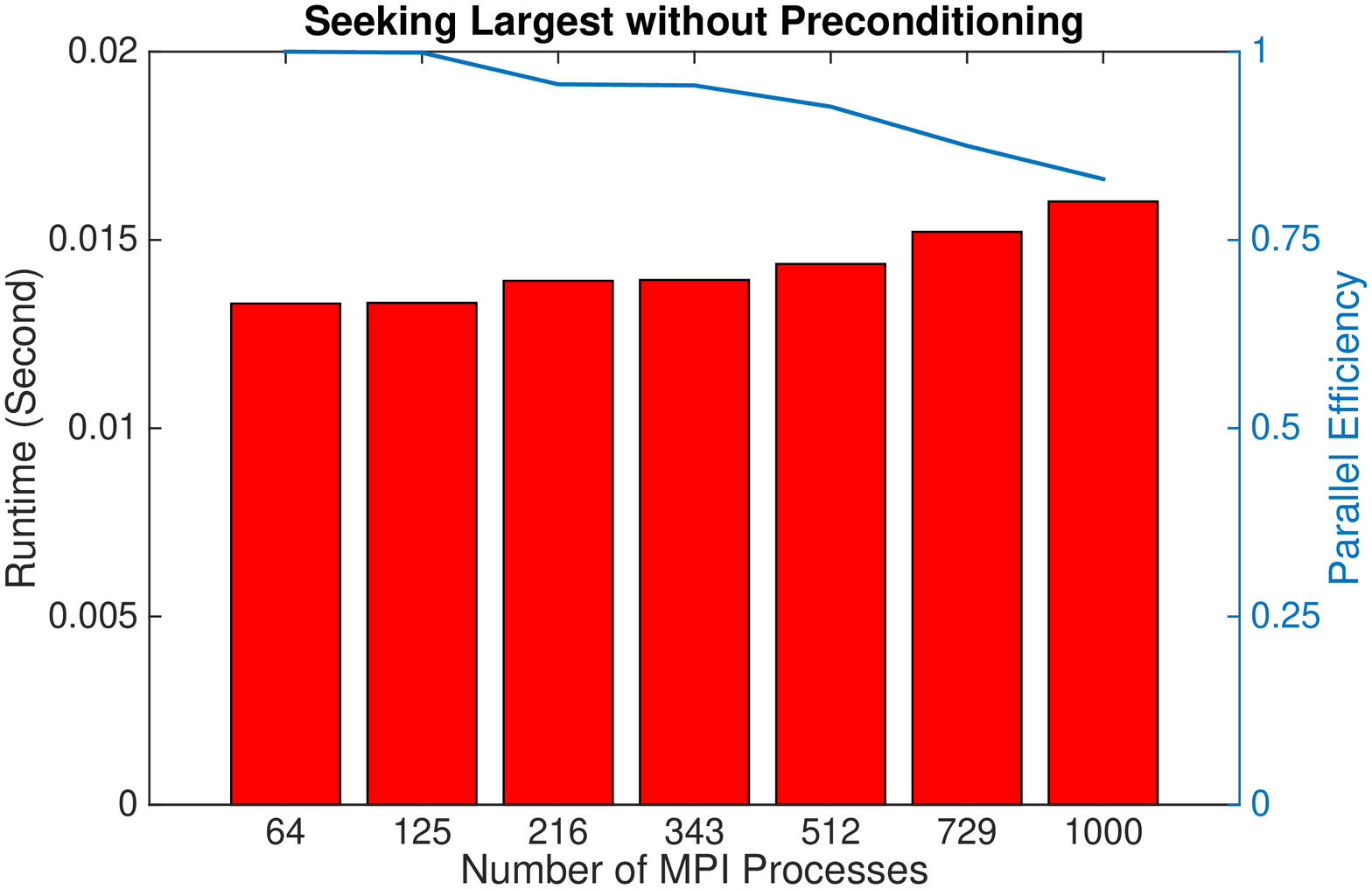}
         \caption{}
         \label{fig:WeakScaling_laplace_LA}
     \end{subfigure}
     \caption{Speedup and runtime seeking the 10 smallest singular triplets in
	     (a) relat9 without preconditioning, 
	     (b) cage15 with HYPRE BoomerAMG as preconditioner.
	     (c) Seeking the 10 largest in delaunay\_n24.
	     (d) Parallel efficiency and runtime seeking the 50 largest 
     		of a 3D Laplacian with fixed 8,000 rows per MPI process.}
     \label{fig:StrongScaling_WeakScaling}
\end{figure}

To further study the parallel scalability of the code, we again use the 
delaunay\_n24 sparse matrix and also the two somewhat denser matrices,
cage15 and relat9. The relat9 is rectangular with a small dimension of 
about half a million which is used as a stress test for strong scalability.
We also test the weak parallel scalability of the code using a series of 
3D Laplacian matrices, making one of its dimensions proportional to the 
number of processes; each process maintains 8,000 rows when the number
of the MPI processes increases from 64 to 1000. 
The plots in Figure \ref{fig:StrongScaling_WeakScaling} show the 
scalability performance of PRIMME\_SVDS on Edison when seeking 10 extreme 
singular triplets with and without preconditioning. 

In Figure \ref{fig:StrongScaling_rectangular_SA}, PRIMME\_SVDS can achieve 
near-ideal speedup until 256 processes on relat9, despite the small size.
With 512 processes, the speedup starts to level off as each process 
has only about 1,000 rows of a very sparse matrix.
In Figure \ref{fig:StrongScaling_square_SA_Prec}, we use the HYPRE BoomerAMG
  multigrid preconditioner so the parallel efficiency is dominated by this 
  library function.
Still, the speedup is good up to 512 processes implying that good 
  preconditioners should be used when available.
Figure \ref{fig:StrongScaling_square_LA} illustrates the same good scalability
performance when seeking largest singular triplets without preconditioning.
In Figure \ref{fig:WeakScaling_laplace_LA} the code demonstrates good performance 
  under weak scaling of the 3D Laplacian on a cubic grid, where each dimension is
  $p^{\frac 1 3}\times 20$ and the number of processes $p$ takes the values of all
  perfect cubes from 64 to 1000.

\subsection{Approximating the condition number}

A particular strength of our software is in seeking one extreme singular value.
This has many important applications such as in computing pseudospectra,
  the matrix norm, or the more difficult problem of approximating 
  the condition number.
A challenge arises, however, when these quantities are needed only 
  in very low accuracy.
Any Davidson or Krylov type method may miss the extreme eigenpair (i.e., 
  misconverge to an interior one) if low accuracy is needed and 
  the spectrum is clustered close to the target eigenvalue. 
A common approach is to use a large enough block size, but there is a 
  performance penalty especially if the rest of the eigenvalues in the block
  are not needed.
For the matrices we test, a block size of one and stopping the
eigensolver when the residual norm is ten times smaller than the approximate
eigenvalue suffices to obtain the smallest singular value with a relative error
of 10\%.

To the best of our knowledge, the most recent high-performance package
  for approximating the condition number is based on solving a linear 
  least-squares problem with a known solution using LSQR \cite{2013arXiv1301.1107A}. 
Its implementation in 
libskylark\footnote{\url{http://xdata-skylark.github.io/libskylark/}}
first approximates the largest singular value with 300 iterations of power method on $C$.
For the smallest, the authors proposed several stopping criteria but for all
cases we test their method stopped when the relative forward error of the 
solution vector $\|\bx-\tbx\|/\|\bx\|$ is less than
$\sqrt{2}\,\mathrm{erf}^{-1}(10^{-3})$. The authors refer to this as small-error
stopping criterion, which they claim may fail with a small probability.
It is unclear what the corresponding probability of failure of our criterion is.

Table~\ref{ta:cond} lists the aggregate matrix-vector products and time for 
computing one largest and one smallest singular value on a single SciClone node 
(12 cores) by PRIMME\_SVDS using the librsb matrix-vector product and libskylark 
using the Elemental shared memory sparse matrix-vector product. 
In general, LSQR requires fewer matrix-vector products than PRIMME\_SVDS
without preconditioning, and achieves a smaller relative error of the condition 
number by at least a factor of ten. 
This is because of the underlying unrestarted LBD method. 
However, PRIMME\_SVDS is much faster in time for a variety of reasons. 
First, the matrix-vector product of librsb was on average two times faster than 
  the Elemental one.
Second, and more importantly, for tall skinny rectangular matrices PRIMME\_SVDS
  works on $C$ which is of much smaller dimension than the full size problem 
  in LSQR.
Finally, libskylark uses 300 power iterations to find the largest 
  singular value which is often unnecessary.
In addition, our software is far more flexible as it can compute the condition
  number with any required accuracy, not only at the tested 10\%, and it can 
  use preconditioning.
Table~\ref{ta:cond} shows the enormous reduction in time resulting from 
   using the preconditioners of the previous sections.


\begin{table}[t]
	\caption{Total matrix-vector products and runtime for 
		PRIMME\_SVDS and \code{CondEst} function in libskylark
		seeking the largest and the smallest singular triplet 
		to estimate the condition number. 
         PRIMME\_SVDS stops at relative tolerance $0.1$. With preconditioning
 	it uses GD+k for the first stage, and without it uses JDQMRETol.}
   \label{ta:cond}
   \small
   \centering
   \newcommand{\rheader}[1]{\multicolumn{2}{c}{#1}}
   \newcommand{\B}[1]{\textbf{#1}}
   \newcommand{\E}[1]{\emph{#1}}
   \begin{tabular}{crrrrrrrr}
           & \rheader{PRIMME\_SVDS} & \rheader{libSkylark}     & \rheader{PRIMME\_SVDS} \\
           & \rheader{(JDQMRETol)}  & \rheader{\code{CondEst}} & \rheader{(GD+k) Prec.} \\
      Matrix         & \multicolumn{1}{c}{MV} & \multicolumn{1}{c}{Sec}&
                       \multicolumn{1}{c}{MV} & \multicolumn{1}{c}{Sec}&
                       \multicolumn{1}{c}{MV} & \multicolumn{1}{c}{Sec}\\\hline
      cage15         & \B{  120}&\B{   53} &     816 &    400  &    25 &   10 \\
      atmosmodl      &    31230 &    2278  &\B{21216}&\B{1540} &    38 &   33 \\
      Rucci1         & \B{36410}&\B{  884} &   64083 &   2860  &  1259 &   92 \\
      LargeRegFile   &     5764 &\B{  217} &\B{ 4470}&    306  &    75 &    6 \\
      sls            &     4172 &\B{   91} &\B{ 3144}&    192  &    24 &    2 \\ \hline
\end{tabular} 
\end{table}

\section{Conclusion and Future Work}
PRIMME\_SVDS is a high-performance package for the computation of a
small number of singular triplets of large-scale matrices. Currently it
includes an implementation of the state-of-the-art PHSVDS method, which
expertly combines the solution of two equivalent eigenvalue formulations to
obtain both performance and accuracy. Previously, the strategy has been 
compared favorably with other state-of-the-art methods, and in this paper
we showed that the PRIMME\_SVDS implementation improves over the best current
SVD packages of SLEPc and PROPACK.

We have discussed the critical aspects of the implementation of the
solver that affect robustness and performance, such as the extraction method
and the heuristic to switch between the normal equations and the augmented
approach. Furthermore we have described the interface, highlighting the
considerations for parallel execution, and illustrated the usage of the package
with a few examples. Finally, we include numerical experiments with problems with
large condition numbers and packed spectrum that complement previously
published experiments to show the clear advantages of PRIMME\_SVDS.

The software package is released as part of PRIMME version 2.1 which is
freely available at \url{https://github.com/primme/primme}. 
Future extensions include a JDSVD method which can also be tuned 
  for use in the second stage of PHSVDS.


\section*{Acknowledgment}
The authors are indebted to the two referees for their meticulous reading.
They would also like to thank the Scientific Data Management Group for
generously help on running experiments on Edison. This work is supported by NSF
under grants No. CCF 1218349 and ACI SI2-SSE 1440700, by DOE under a grant No.
DE-FC02-12ER41890 and partially supported by the Office of Advanced Scientific
Computing Research, Office of Science, of the U.S. Department of Energy under
Contract No. DE-AC02-05CH11231. This work was performed in part using
computational facilities at the College of William and Mary which were provided
with the assistance of the National Science Foundation, the Virginia Port
Authority, Sun Microsystems, and Virginia's Commonwealth Technology Research
Fund.

\bibliographystyle{siam}

\bibliography{primme_svds2}

\end{document}